\documentclass[twocolumn,trackchanges]{aastex62}

\graphicspath{{./}{figures/}}
\usepackage{multirow,,color,txfonts}
\usepackage{makecell}

\begin{document}

\title{A systematic study of $\gamma$-ray flares from the Crab Nebula
with Fermi-LAT: I. flare detection}

\author{Xiaoyuan Huang$^1$, Qiang Yuan$^{1,2,3}$, Yi-Zhong Fan$^{1,2}$}
\affil{$^1$Key Laboratory of Dark Matter and Space Astronomy, Purple Mountain 
Observatory, Chinese Academy of Sciences, Nanjing 210033, China\\
$^2$School of Astronomy and Space Science, University of Science and 
Technology of China, Hefei, Anhui 230026, China\\
$^3$Center for High Energy Physics, Peking University, Beijing 100871, China}
\email{xyhuang@pmo.ac.cn (XH), yuanq@pmo.ac.cn (QY)}

\begin{abstract}
Significant flares of GeV $\gamma$-ray emission from the Crab Nebula have
been found by AGILE and Fermi-LAT years ago, indicating that extreme
particle acceleration and radiation occurs in young pulsar wind nebulae.
To enlarge the flare sample and to investigate their statistical
properties will be very useful in understanding the nature of the
$\gamma$-ray flares. In this paper, we investigate the flaring emission 
from the Crab Nebula with eleven year observations of the Fermi-LAT. 
We identify 17 significant flares in the light curve of the low-energy 
(synchrotron) component of the $\gamma$-ray emission. The flare rate
is about 1.5 per year, without any significant change or clustering
during the 11 years of the observation. We detect a special flare with an extremely long duration of nearly
one month, occurred in October, 2018, with synchrotron photons up to energies of about 1GeV. The synchrotron component could be fitted by a
steady power-law background and a variable flare component with an expotentially
cutoff power-law spectrum, not only for individual flare but also for the combined data, which may favor a similar emission mechanism for all flares. However, we do not find a universal relation between the cutoff energy and the energy fluxes of the flares, which may reflect the complicated acceleration and/or cooling processes of the involved particles. 
\end{abstract}


\section{Introduction} \label{sec:intro}

The Crab Nebula, one of the most interesting and well studied objects 
in the sky, is powered by a young and energetic pulsar, PSR B0531+21, 
which is a remnant of the supernova recorded by Chinese astronomers in 
AD 1054. A wind of cold ultra-relativistic particles, accelerated by 
the rapidly rotating, powerful magnetic fields of the central pulsar, 
terminates where its momentum flux density is balanced by the confining 
pressure of the external medium (for young pulsars it may be the ejected
stellar material, and for old pulsars it is the interstellar medium). 
The forming termination shock accelerate high-energy electrons, which
lighten the nebula in wide wavebands \citep{1984ApJ...283..694K, 1994ApJS...90..797A}. 
The very broad spectrum of the Crab Nebula can be largely attributed to 
the synchrotron radiation by relativistic electrons with energies from 
GeV to PeV \citep{CoCKE:1970CD,1972ApJ...174L...1N,2008Sci...321.1183D}, 
and the inverse Compton (IC) scattering emission off the cosmic microwave
background, the synchrotron nebula, and the thermal dust emission 
\citep{1965AnAp...28..171G,1996MNRAS.278..525A,2010A&A...523A...2M}. 

For a long time, the overall emission from the Crab Nebula was expected to be 
steady. The Crab Nebula is regarded as a ``standard candle'' and is usually 
used to cross-calibrate X-ray and very high energy $\gamma$-ray telescopes 
\citep{Kirsch:2005tn,2010ApJ...713..912W,2010A&A...523A...2M}. 
However, the MeV $\gamma$-ray emission seems not that simple. Early 
observations by COMPTEL and EGRET already showed possible flux variations 
at a time scale of one year \citep{1995A&A...299..435M,1996ApJ...457..253D}.
The overall flux in the hard X-ray band also shows shallow changes of 
a few percents in several years \citep{2011ApJ...727L..40W}.
Surprisingly strong flares have been found for energies above 100 MeV, by 
AGILE \citep{2011Sci...331..736T} and Fermi-LAT \citep{2011Sci...331..739A}.
Later several more flares have been reported \citep{2011ApJ...741L...5S,
2012ApJ...749...26B,2013ApJ...775L..37M,2013ApJ...765...52S}, with a 
super-giant one occurred in April, 2011. Dedicated efforts have been 
paid to search for possible counterparts of the $\gamma$-ray flares in 
other wavelengths, but no firm counterpart has been found yet \citep{2011A&A...533A..10L,2012ATel.4258....1B, 2013ApJ...765...56W, 2014RPPh...77f6901B, 2014A&A...562L...4H,2014ApJ...781L..11A,2015JHEAp...5...30A,2015ApJ...811...24R}.

Various models have been proposed to explain the $\gamma$-ray flares,
with particular focus on the puzzle that how could electrons generate 
synchrotron radiation above a maximum energy of $\sim 160$ MeV for the
classical shock acceleration \citep{1983MNRAS.205..593G,2011ApJ...737L..40U}, 
and what is the location of the emission sites where rapid variability on 
a timescale of hours could be produced. A widely adopted way to produce
synchrotron emission up to GeV energies is the Doppler boosting of the
emission site \citep{1996MNRAS.283L.133C, 2011MNRAS.414.2017K,2011ApJ...730L..15Y,
2012MNRAS.424.2249K}. Alternatively, the magnetic reconnection induces 
a linear electric accelerator which can also overcome such a difficulty
\citep{2011ApJ...737L..40U,2012ApJ...746..148C,2012ApJ...754L..33C,
2013ApJ...770..147C,2014PhPl...21e6501C,2016ApJ...828...92Y,
2017ApJ...847...57Z,2017JPlPh..83f6301L,2017JPlPh..83f6302L}.
As for the flare site, in \citet{2011MNRAS.414.2017K} it was proposed 
that the observed synchrotron $\gamma$-rays would be dominated by the 
contribution of the inner knot, whose size is about a few light days
and is consistent with the flare duration. Also the emissions from the 
inner knot would be blue-shifted and can exceed the $160$ MeV limit. 
Rapid changes of the shock geometry due to the violent dynamics of the 
inner nebula may produce the observed variability 
\citep{2011MNRAS.414.2017K,2012MNRAS.422.3118L}, with a possible caveat
that no correlated variability in various energy bands was observed 
\citep{2015ApJ...811...24R}. \citet{2011MNRAS.414.2229B} suggested that 
electrons are accelerated in a region behind the shock, and the variability 
is attributed to changes in the maximum energy of accelerated electrons, 
electron spectral index, or the magnetic field. This scenario predicts
multi-TeV $\gamma$-ray variabilities as a result of the IC emission which
is lacking. \citet{2017PhRvL.119u1101K} proposed that the frequency, 
variability and power of the flares emerge as natural consequences of a 
sharp reduction of the supply of electron-positron pairs to the wind of 
the Crab pulsar, furthermore the polarization properties of the flares 
and possible similar emission from other pulsar wind nebulae are predicted. 

Previous works for the analysis of the $\gamma$-ray flares from the Crab 
Nebula were generally based on case studies \citep{2011Sci...331..736T, 
2011Sci...331..739A,2012ApJ...749...26B,2011ApJ...741L...5S,
2013ApJ...775L..37M}. Given the long-term operation of Fermi-LAT for more
than 10 years, it is expected that more $\gamma$-ray flares would be
detected, and it is highly desired to have a population study of the flares.
This is the motivation of the current study. The statistical characterization
of the flare properties is expected to be very useful in revealing the
physical nature of the flares \citep{2016MNRAS.456.1438Y,2018MNRAS.473..306Y}, 
which will be studied in details in an accompanying work. 
The structure of this paper is as follows. 
In Section \ref{sec:phase} we calculate the folded pulsar phase of each 
photon based on radio observations in order to remove the strong pulsar
emission. We then extract $\gamma$-ray flares from 11 years of observations 
of the Fermi-LAT in Section \ref{sec:flares}, and carry out a case study for the ultra-long 
duration flare occurred in October, 2018, in Section \ref{sec:2018}. 
In Section \ref{sec:combined} we investigate briefly the
properties of all detected flares. We discuss the possible
implications of our results in Section \ref{sec:discussion}, and then 
conclude our study in Section \ref{sec:conclusion}.

\section{Phase folding of the central pulsar}\label{sec:phase}

The Crab pulsar, PSR B0531+21, one of the most energetic known pulsars 
(with a spin down power of $\dot{E}=4.6\times10^{38}$~erg\,s$^{-1}$), 
lies at the center of the Crab Nebula. To remove the strong $\gamma$-ray 
emission from this pulsar, we need to calculate the phase of each photon 
and to select photons in the off-pulse window for the following analysis 
about the nebula.

The rotation rate of the Crab pulsar, like many young pulsars, is affected 
by significant timing noise and glitches. Since we will cover a relatively 
long time interval in this paper, the rotational behavior evolution with 
time needs to be known with a very high precision. The Jodrell Bank 
Observatory has continuously made the monthly ephemeris of the Crab pulsar 
for decades\footnote{http://www.jb.man.ac.uk/pulsar/crab.html} 
\citep{1993MNRAS.265.1003L}. These monthly ephemeris data give the primary 
spin frequency (F0), the first derivative (F1), the rate at which the pulsar 
slows down, and also the second derivative (F2) which gives the timing noise, 
of the Crab pulsar for time windows lasting about one month each. With these 
ephemeris data we can use the \textit{TEMPO2}\footnote{https://www.atnf.csiro.au/research/pulsar/tempo2/}, a pulsar analysis package developed by radio 
astronomers, to assign phases to $\gamma$-ray photons collected by the 
Fermi-LAT with the help of a plugin\footnote{http://fermi.gsfc.nasa.gov/ssc/data/analysis/user/Fermi$\_$plug$\_$doc.pdf}. 
We also assume that the rotation of the Crab pulsar will not change 
significantly in a short time. Therefore for photons with arrival times
outside a given time window as defined in the monthly JODRELL BANK CRAB PULSAR MONTHLY EPHEMERIS, it is still fine to calculate the phase 
using the nearby ephemeris. A match of the phases calculated using the
previous ephemeris and the successive one has been applied. 

\begin{figure}
\includegraphics[width=0.48\textwidth]{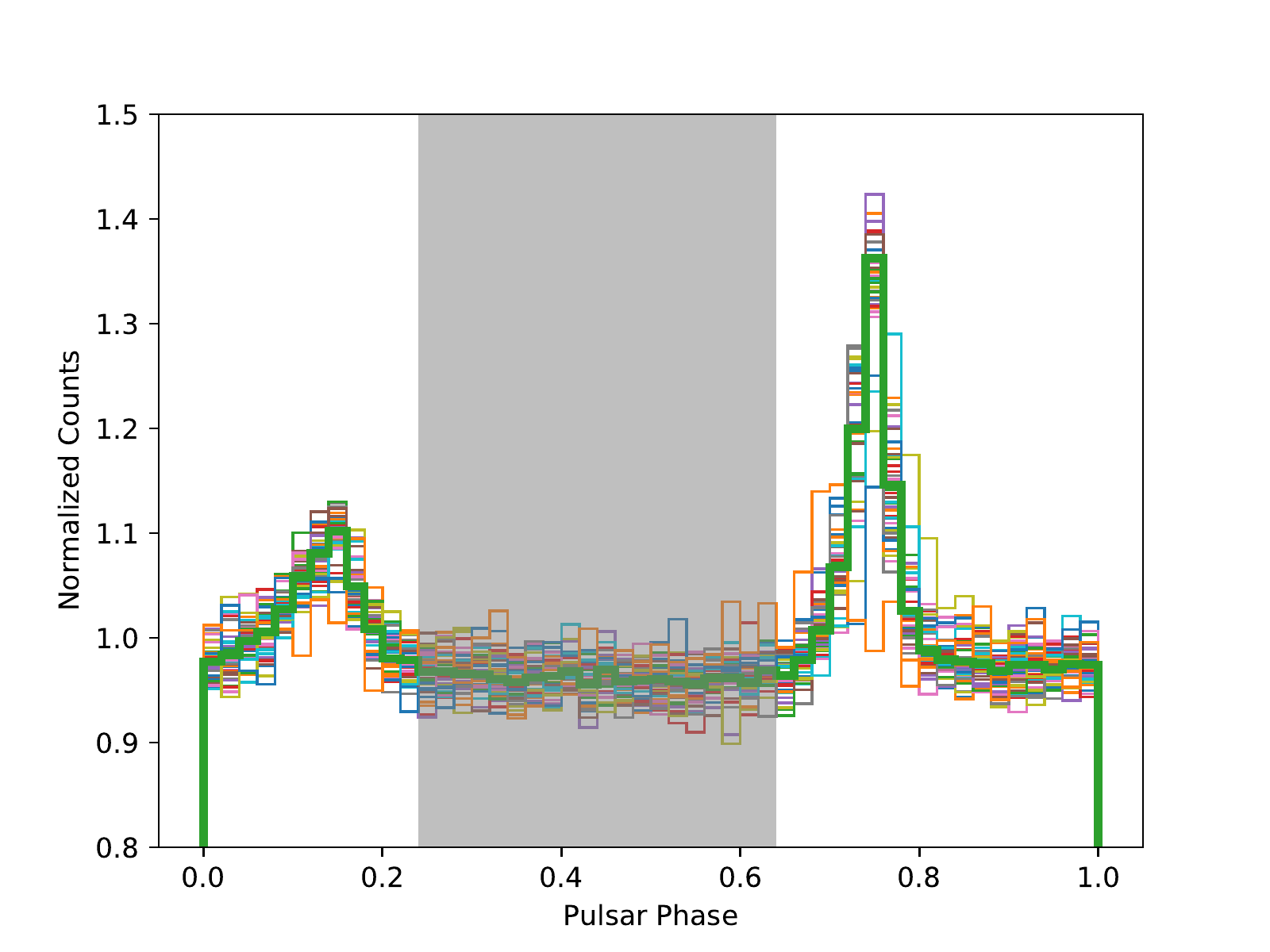}
\includegraphics[width=0.48\textwidth]{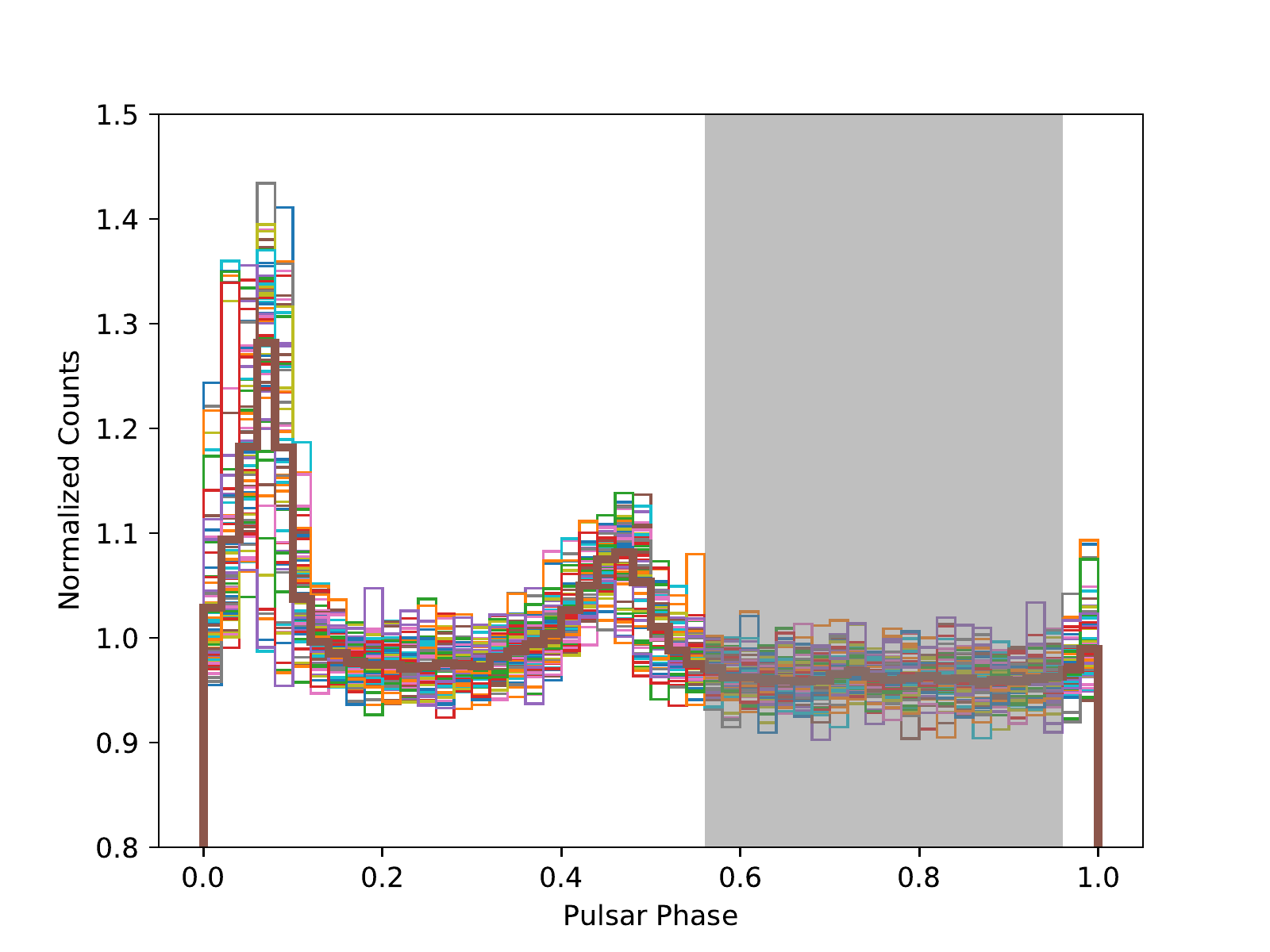}
\includegraphics[width=0.48\textwidth]{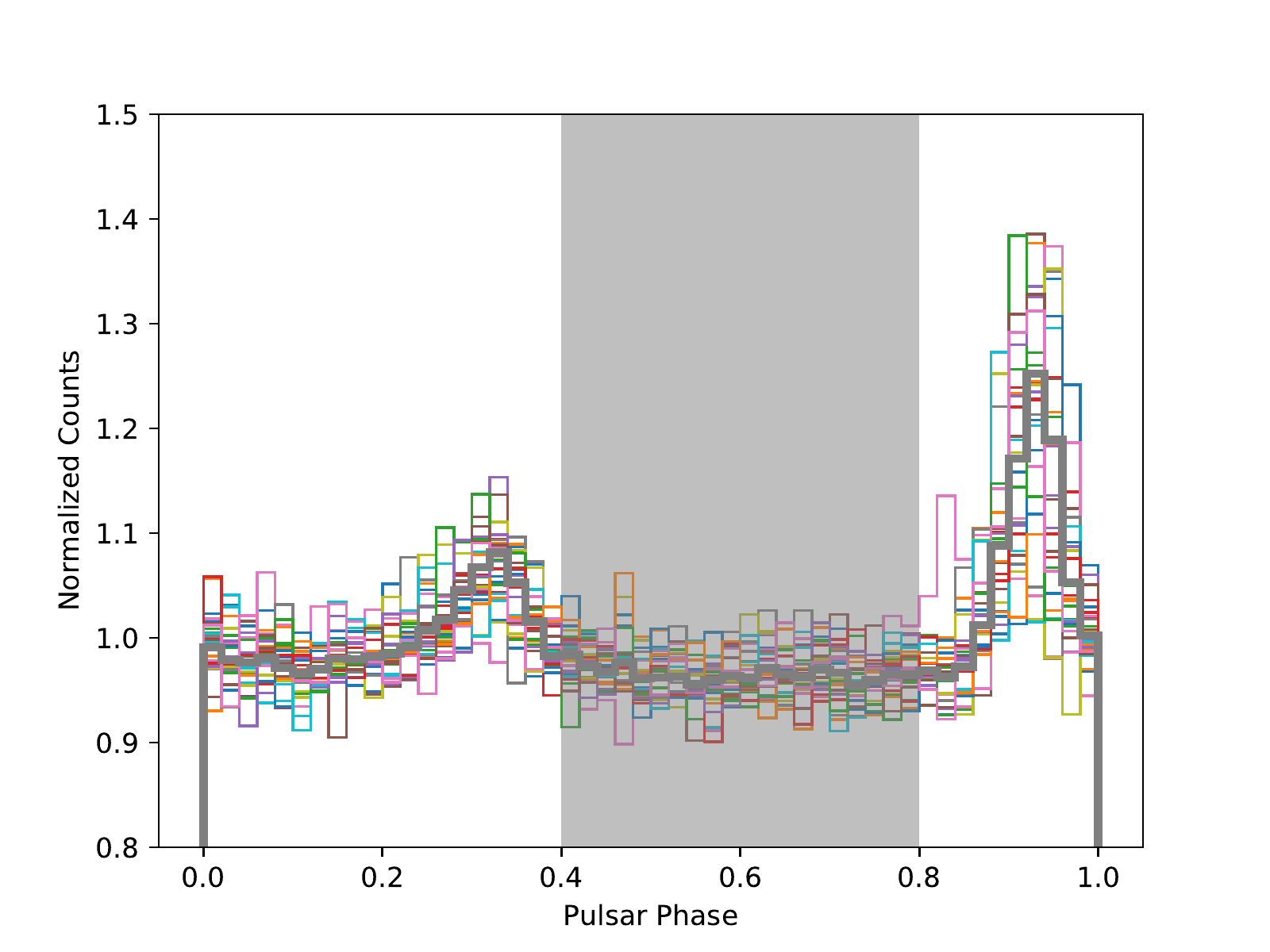}
\caption{The phase-folded light curves in three time intervals, from 
MJD 54682 to MJD 55876 (top), from MJD 55876 to MJD 58063 (middle), 
and from MJD 58065 to MJD 58788 (bottom), respectively. In each panel, 
the thin histograms show light curves for photons selected in each time 
window defined in the JODRELL BANK CRAB PULSAR MONTHLY EPHEMERIS, and 
thick one shows the combined light curve for all photons in this time
period. The grey region in each panel shows the off-pulse window used 
for the following Crab Nebula analysis.
\label{fig:phase}}
\end{figure}

We use the Fermi-LAT data collected from August 4, 2008 to October 24, 2019.
The events that pass the {\tt{P8R3 SOURCE}} event class selections with
energies from 80 MeV to 300 GeV and angles within 25 degrees from the 
Crab pulsar are selected. Here we select photons with energies down to 
80 MeV and angles within 25 degrees from the target source to get more 
photons to calculate the pulsar phases more accurately. The folded light
curve in pulsar phase for three periods are shown in Fig.~\ref{fig:phase}.
The separation into three time periods are based on the times when two 
significant glitches happened on November 10, 2011 and November 8, 
2017\footnote{http://www.jb.man.ac.uk/pulsar/glitches.html}.

\section{Crab flares from the Fermi-LAT data}\label{sec:flares}

We re-select $\gamma$-ray photons that pass the {\tt{P8R3 SOURCE}} event 
class selection, with energies from 100 MeV to 300 GeV and angular 
deviations within 15 degrees from the Crab pulsar for the flare analysis.
For the three time intervals defined in Fig.~\ref{fig:phase}, the off-pulse
phase ranges are defined as 0.24 to 0.64, 0.56 to 0.96, and 0.40 to 
0.80\footnote{We also check that using more strict rotational phase cuts 
of 0.29 to 0.59, 0.61 to 0.91, and 0.45 to 0.75, will not change our 
results, as given in Appendix \ref{sec:phasecut}.}, as shown in gray bands 
in Fig. \ref{fig:phase}. Photons collected at zenith angles larger than 
90$^\circ$ are removed to suppress the contamination from $\gamma$-rays 
generated by cosmic-ray interactions in the upper layers of the atmosphere. 
Moreover, we filter the data using the following specifications 
{\tt{(DATA$\_$QUAL$>$0)~$\&\&$~(LAT$\_$CONFIG==1)~$\&\&$~(angsep (83.63, 
22.01, RA$\_$SUN, DEC$\_$SUN)$>$15)}} to select the good time intervals 
in which the satellite is working in the standard data taking mode and 
the data quality is good, and to exclude times when the Crab nebula is 
within 15 degrees of the Sun to suppress the contamination from the Sun's 
activities. We employ the unbinned likelihood analysis method to analyze 
the data with the Fermitools {\tt version 1.2.1}. The instrument response 
function (IRF) adopted is {\tt P8R3\_SOURCE\_V2}. For the diffuse background 
emissions we take the Galactic diffuse model {\tt gll\_iem\_v07.fits} and 
the isotropic background spectrum {\tt iso\_P8R3\_SOURCE\_V2\_v1.txt} as 
recommended by the Fermi-LAT collaboration\footnote{http://fermi.gsfc.nasa.gov/ssc/data/access/lat/BackgroundModels.html}. 
The source model XML file is generated using the user contributed tool 
{\textit {make4FGLxml.py}}\footnote{http://fermi.gsfc.nasa.gov/ssc/data/analysis/user/} based on the 4FGL source catalog \citep{2019arXiv190210045T}. 
Following \citet{2010ApJ...708.1254A}, we assume the emission from the Crab 
Nebula consists of two spectral components, high-energy component for IC emission and low-energy component for synchrotron emission, each with a power-law (PL) spectrum, $dN/dE \propto E^{-\Gamma_{i}}$, taking $i=h, l$ for high and low energy component, respectively.

\begin{figure}
\includegraphics[width=0.48\textwidth]{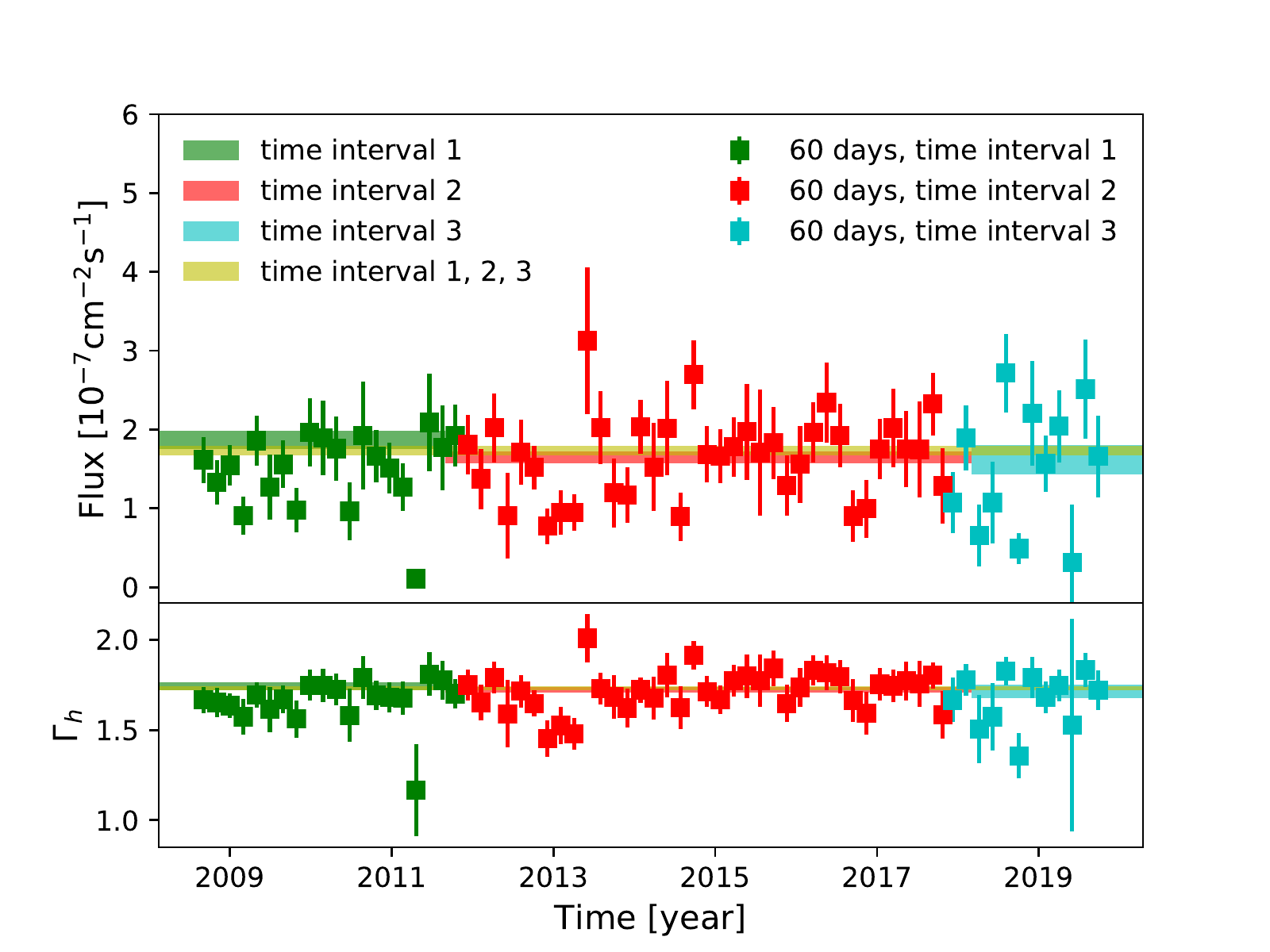}
\includegraphics[width=0.48\textwidth]{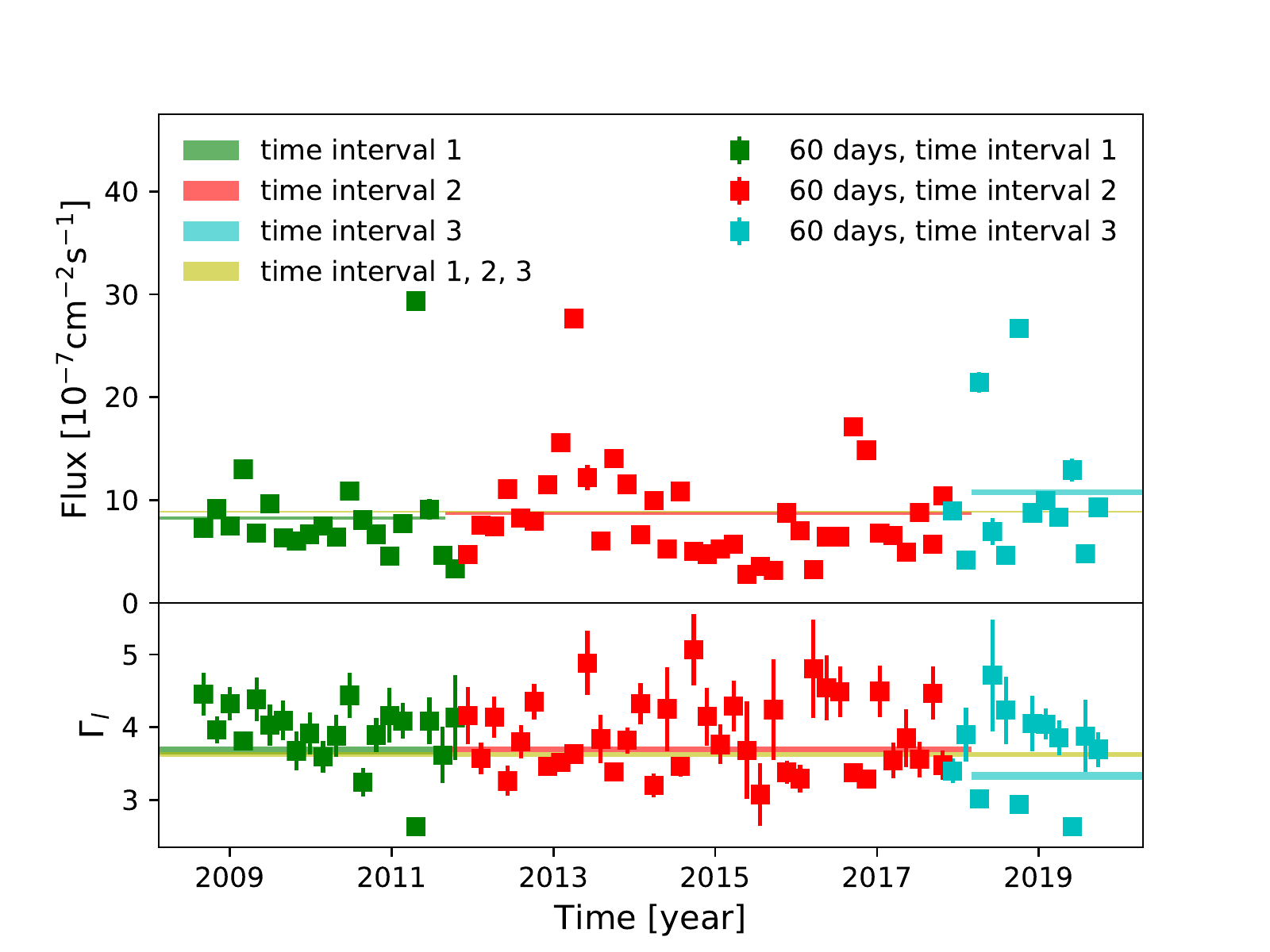}
\caption{Fluxes and spectral indices of the high-energy (IC emission, top panel) and low-energy (synchrotron emission, bottom panel) components 
in 60-day binning. The shaded bands in each panel show the fluxes and spectral indices 
from the likelihood fits done for the three time intervals individually, 
and that from the joint fit.      
\label{fig:flux}}
\end{figure}

We perform an unbinned maximum likelihood fit with the module 
\textit{pyLikelihood} in the Fermitools for each time interval, and use 
the module \textit{SummedLikelihood} to perform a joint likelihood 
analysis of all the three time intervals. For the joint analysis, the 
integrated flux above 100 MeV of the low-energy component is found to be 
$\Phi_{100}=(8.84 \pm 0.08)\times 10^{-7}$~cm$^{-2}$~s$^{-1}$, and the 
photon index is $\Gamma_{l}=3.63 \pm 0.03$, and the high-energy component 
has an integral flux of $\Phi_{100}=(1.73 \pm 0.06)\times10^{-7}$
cm$^{-2}$~s$^{-1}$ and $\Gamma_{h}=1.73 \pm 0.01$. Compared with parameters derived in \citet{2010ApJ...708.1254A,2011Sci...331..739A}, we got a good agreement with spectral indices for those two components, but the integrated fluxes are higher. For each time intervals, 
we also derive the fitting results of these two components. 
As shown in Fig.~\ref{fig:flux} the integrated fluxes for each time 
interval are almost consistent with that from the joint likelihood 
analysis, except for the flux of the low-energy component in the third 
time interval. A higher value of the flux of the low-energy component 
in the third time interval is possibly due to the long-duration flare 
occurred in October, 2018 (see Section \ref{sec:2018}).

Fig.~\ref{fig:flux} shows the light curves and spectral indices of the high-energy (top) and
low-energy (bottom) components for a bin width of 60 days. For each time
bin, the free parameters in the likelihood fit include the normalizations 
and spectral indices of the (both) nebula components, and the normalizations
of the Galactic and isotropic diffuse backgrounds. All other parameters 
are fixed to their best-fit values from the joint likelihood analysis. 
It is clear that the flux of  high-energy component is stable with fluctuations
within $3\sigma$ from the average flux. An only exception is the bin 
around April, 2011\footnote{We find that in this time bin the spectral 
index of the low-energy component, is very hard, 2.63, which is closer 
to the spectral index of the high-energy component. This may make the 
fitting results of these two components degenerate.}. However, the fluxes 
of the low-energy component show significant variations. The highest
flux bins show more than 4 times higher fluxes than the average one,
and have more than $30\sigma$ deviations.

\begin{figure*}
\centering
\includegraphics[width=0.8\textwidth]{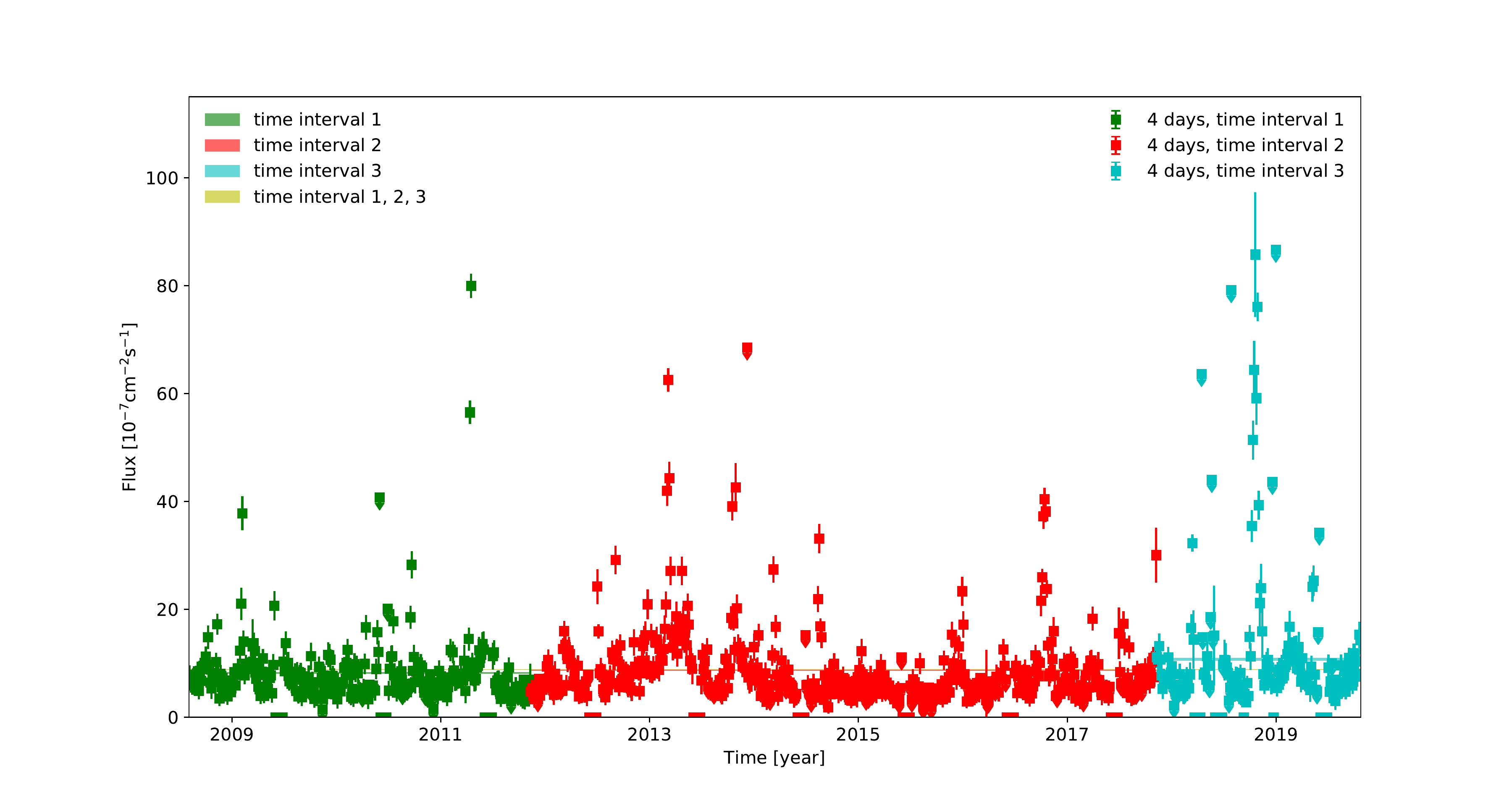}
\caption{Same as the light curves in the  bottom panel of Fig.~\ref{fig:flux} but for 4-day binning.
\label{fig:flux_4days}}
\end{figure*}

To see the varibilities more clearly, we re-bin the data into a bin
width of 4 days\footnote{We chose this width of time bin based on a
balance between the time resolution and photon statistics.}, and re-fit 
the data with only the normalizations and spectral indices of the 
low-energy component being free parameters. The high-energy component 
normalizations and spectral indices and the Galactic and isotropic 
diffuse background normalizations are fixed to their best-fit values 
derived in the joint analysis. The light curve of the low-energy 
component in 4-day time bin is shown in Fig.~\ref{fig:flux_4days}. 

\begin{figure*}
\centering
\includegraphics[width=0.25\textwidth]{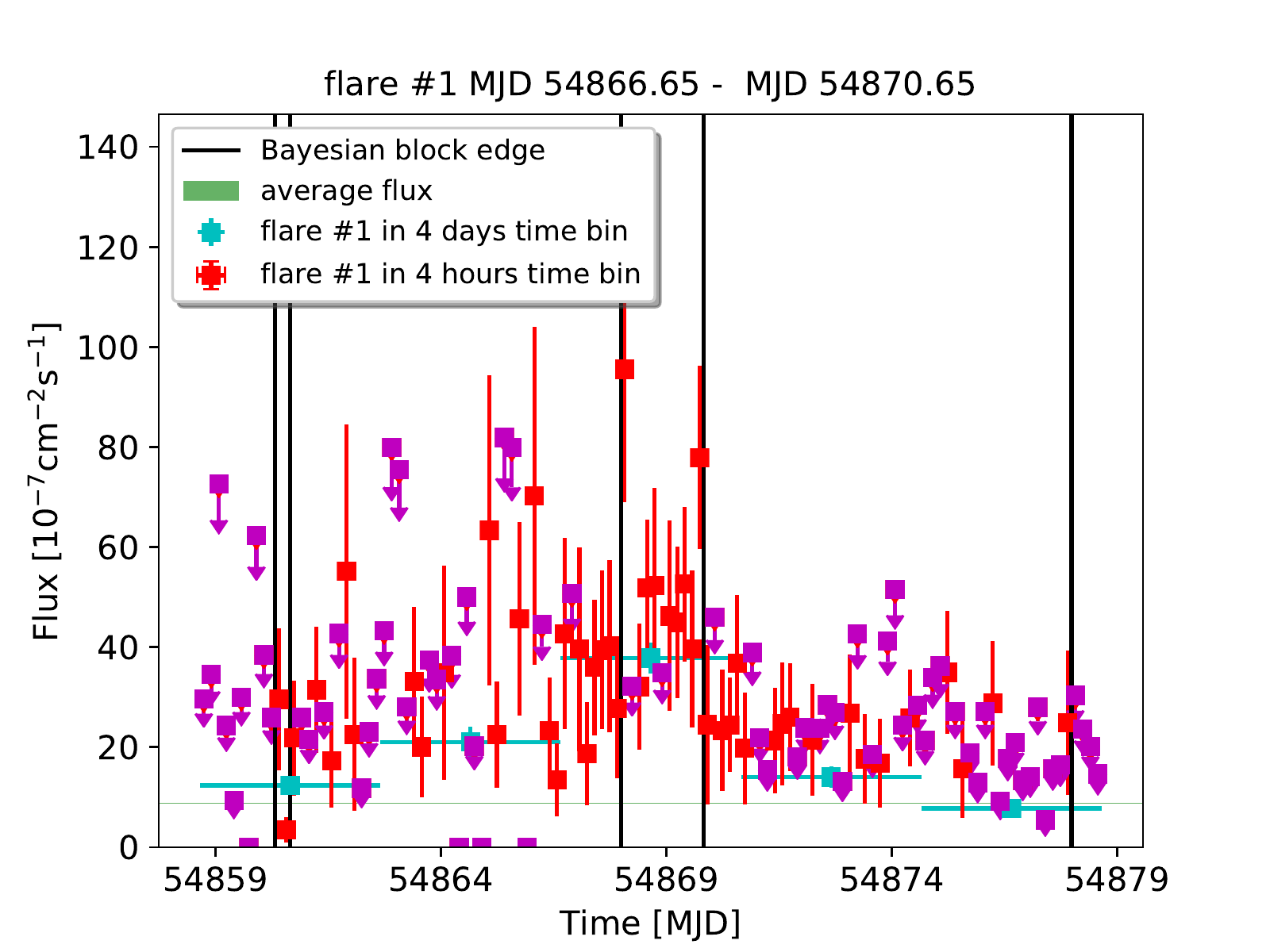}
\includegraphics[width=0.25\textwidth]{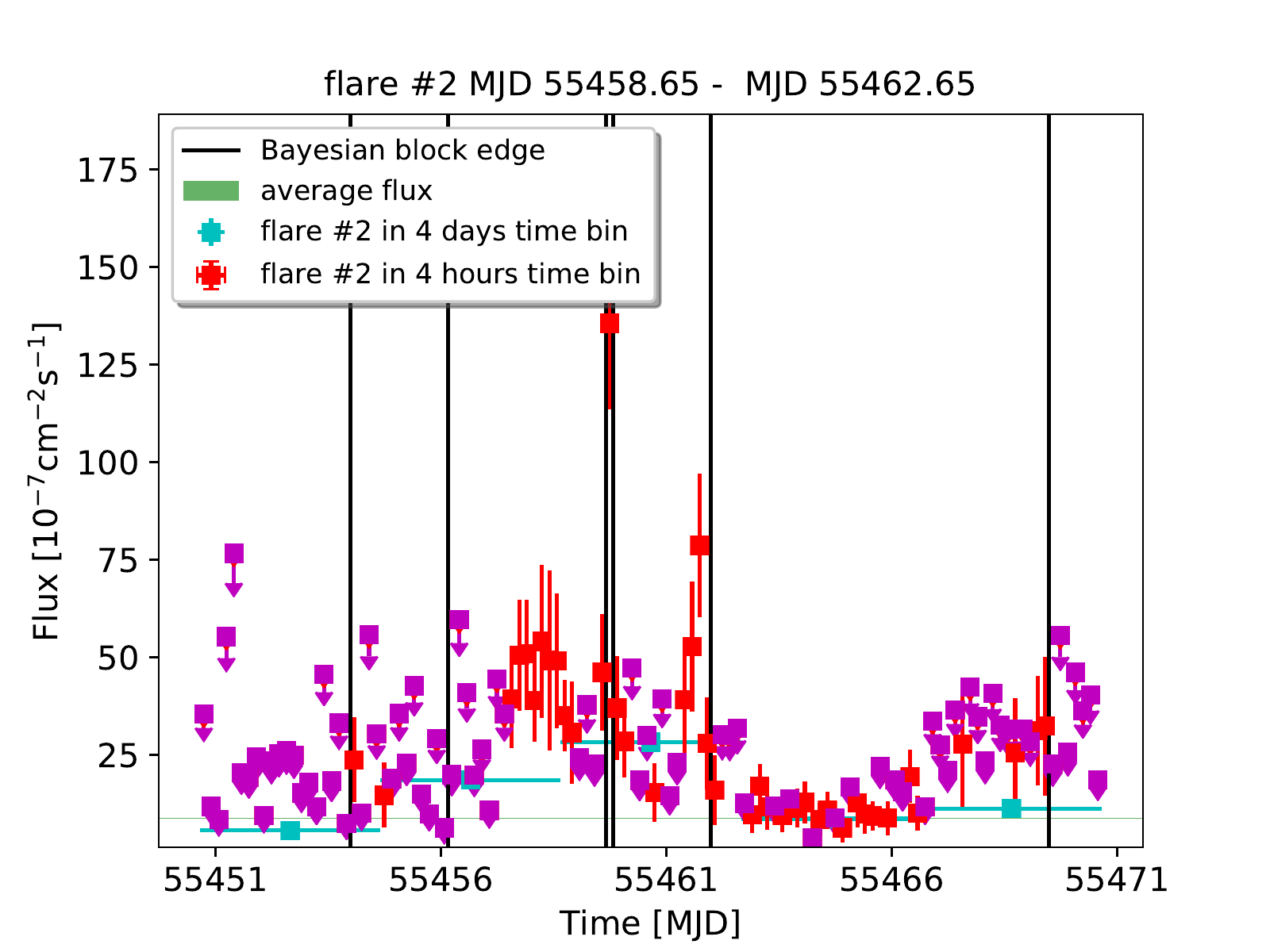}
\includegraphics[width=0.25\textwidth]{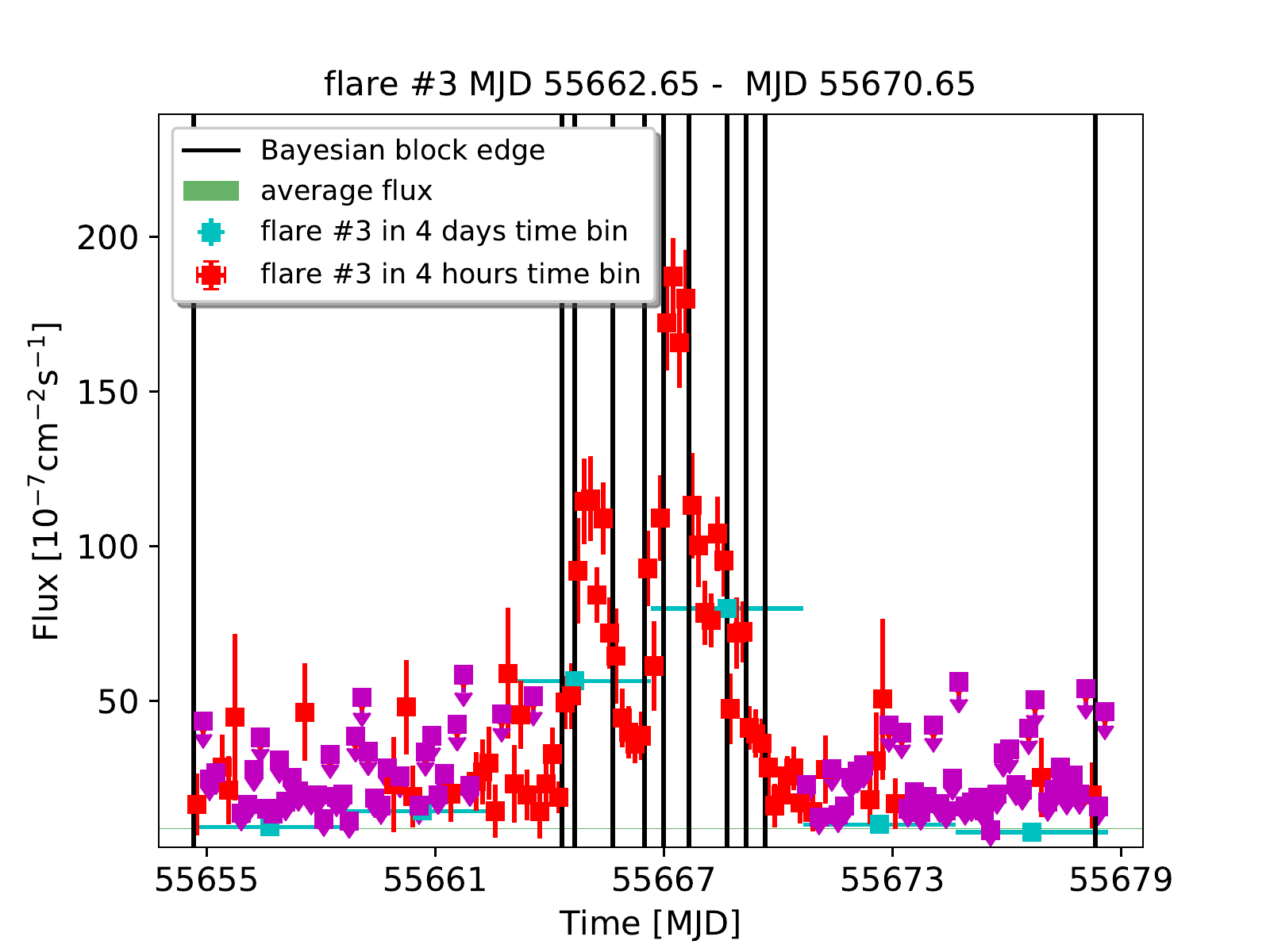}
\includegraphics[width=0.25\textwidth]{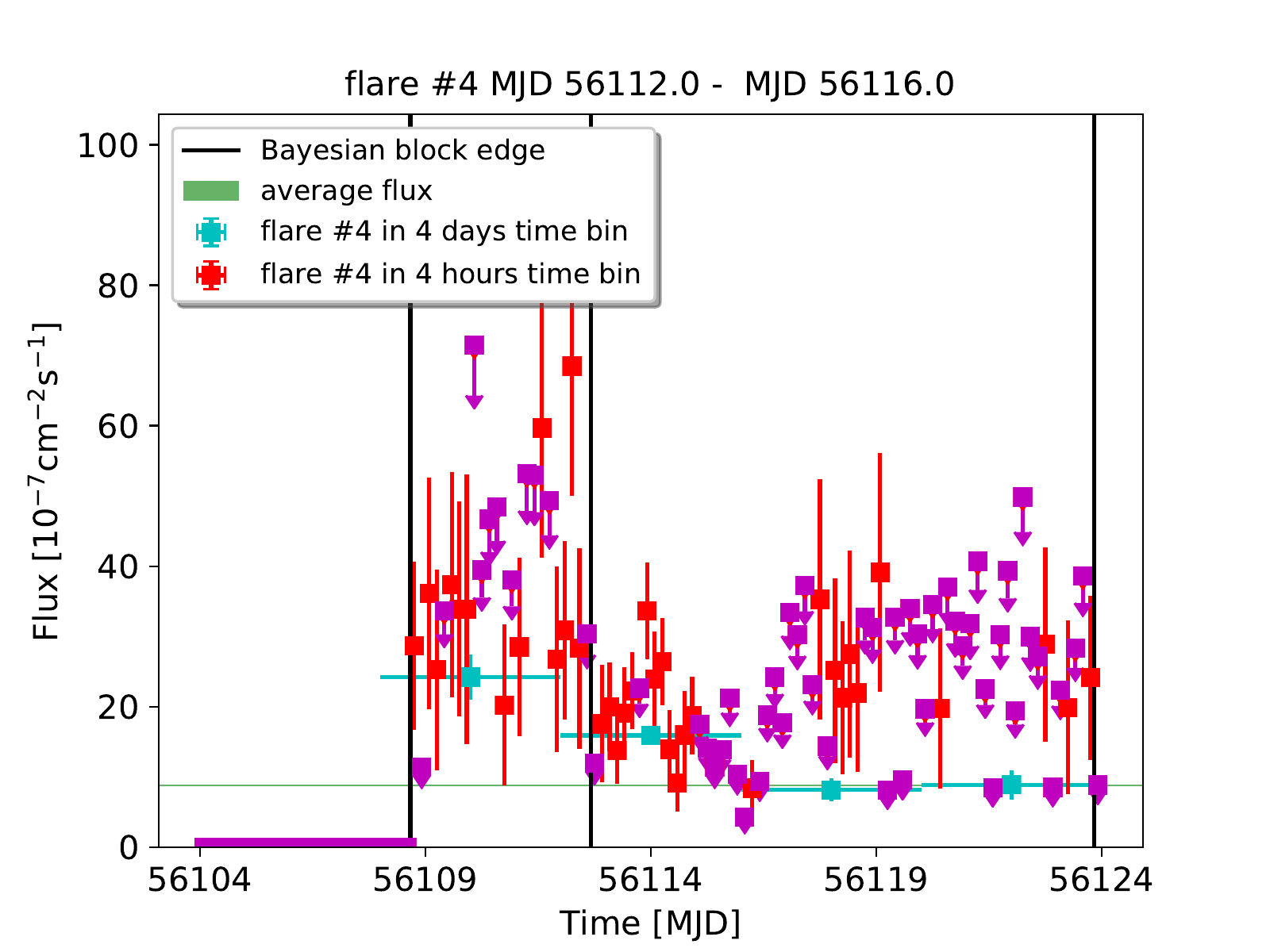}
\includegraphics[width=0.25\textwidth]{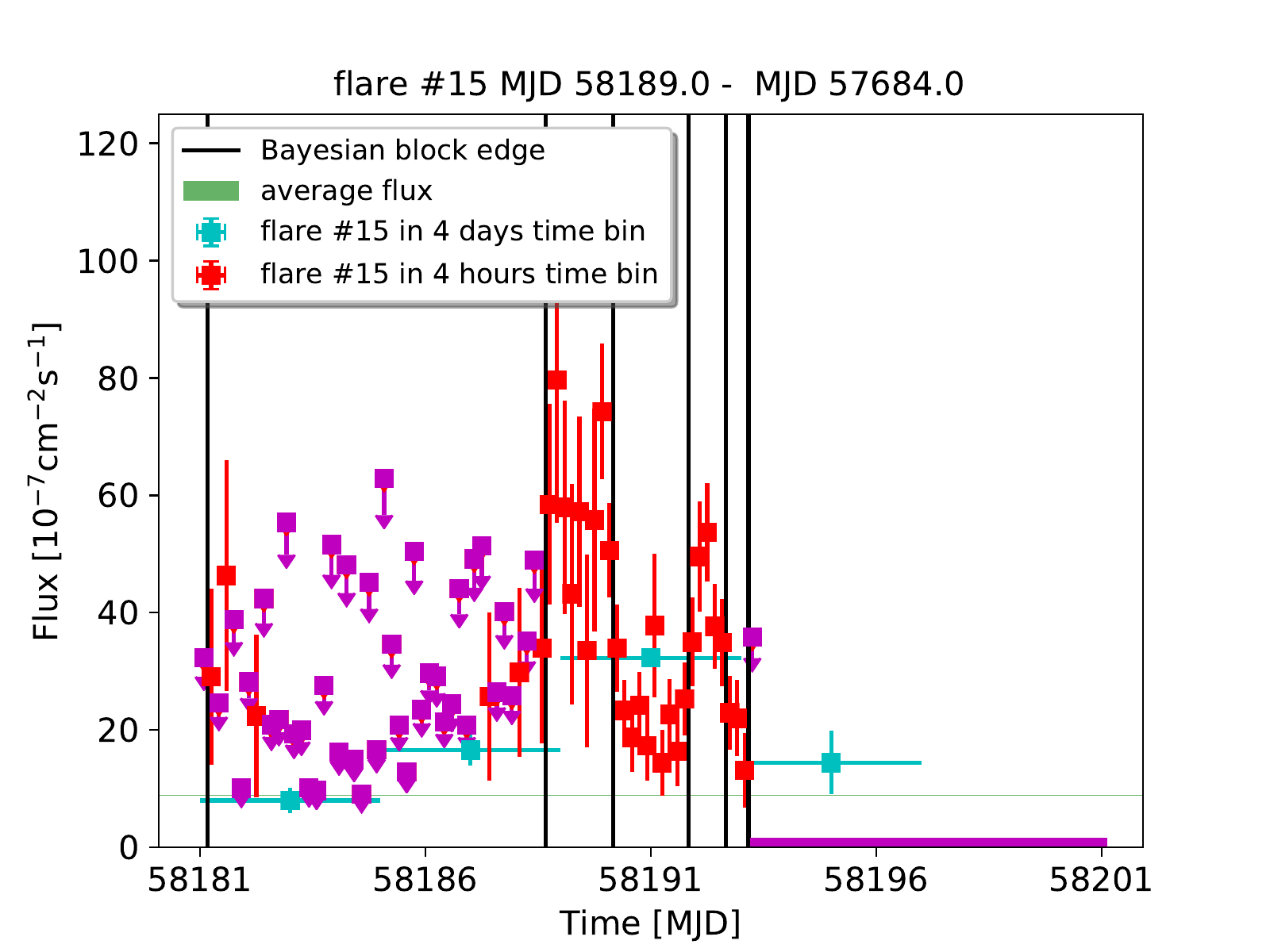}
\includegraphics[width=0.25\textwidth]{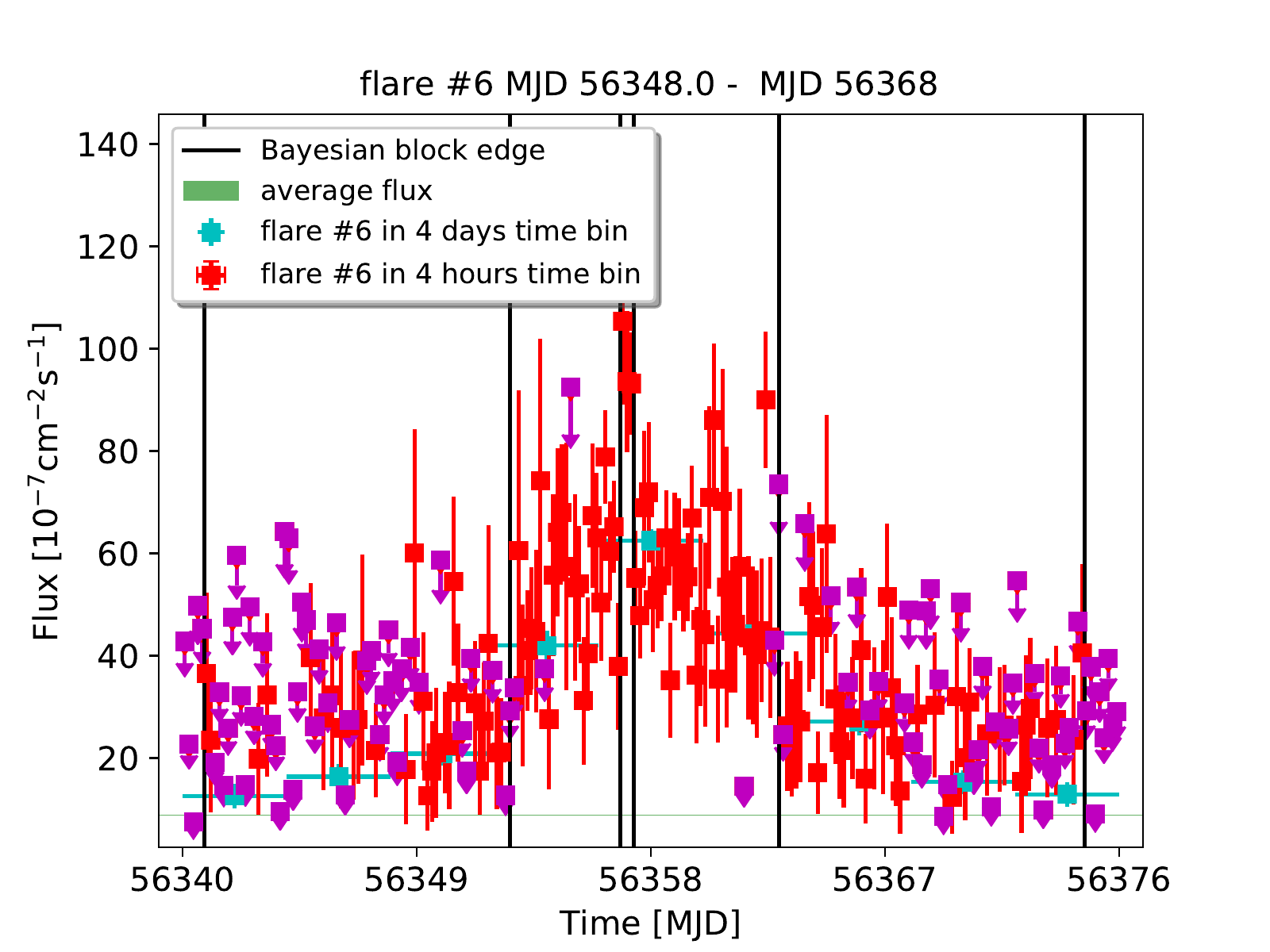}
\includegraphics[width=0.25\textwidth]{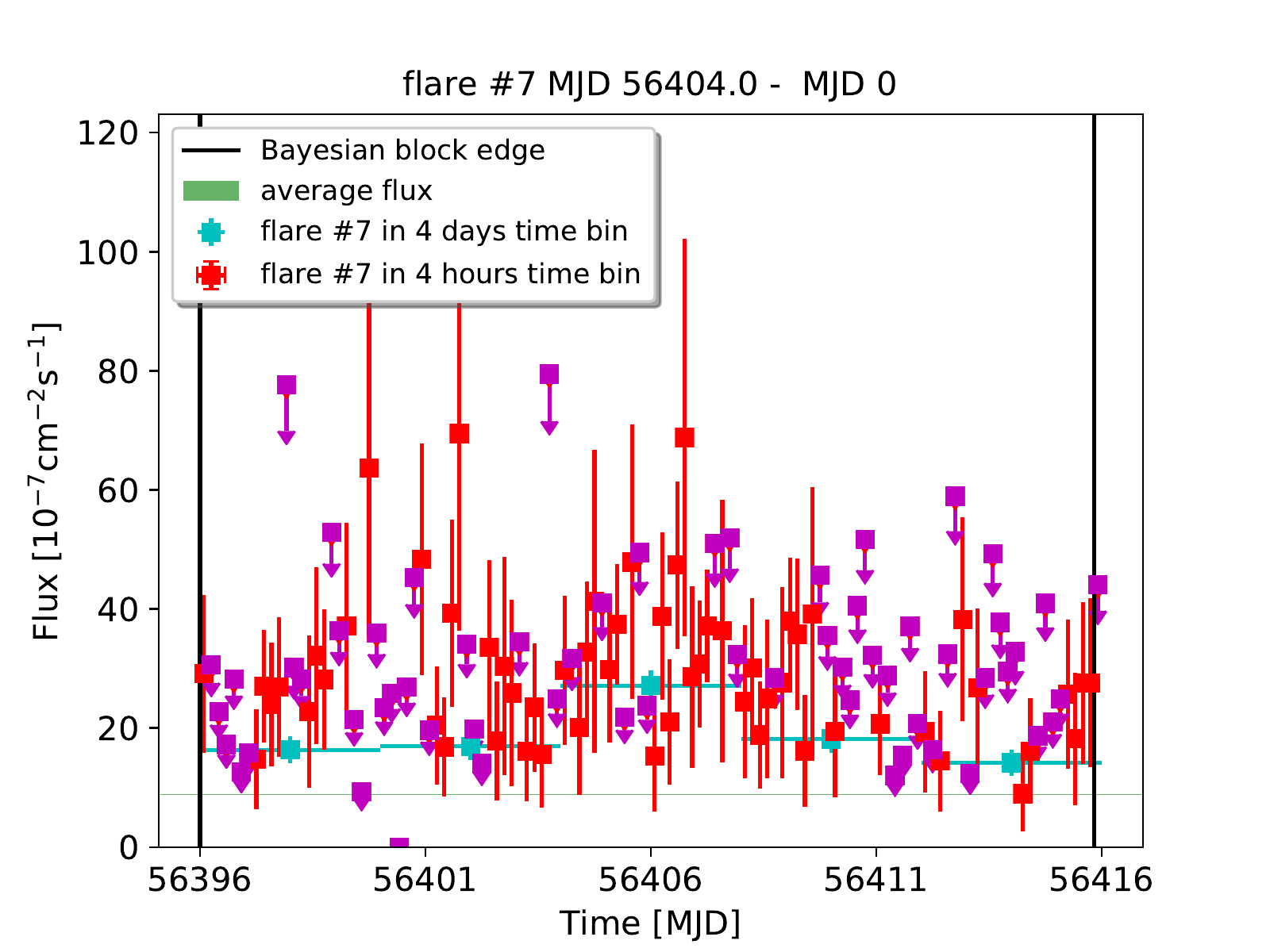}
\includegraphics[width=0.25\textwidth]{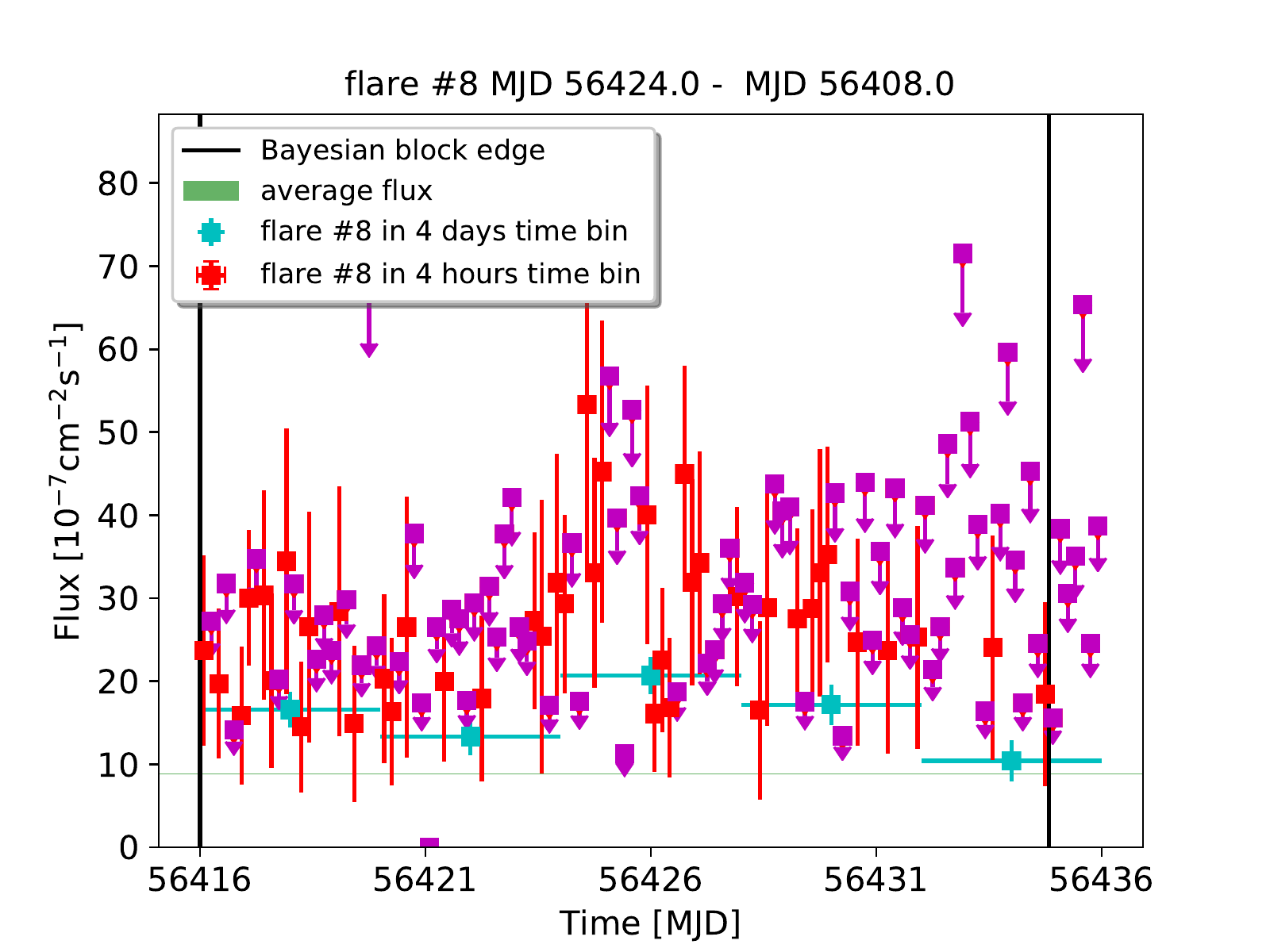}
\includegraphics[width=0.25\textwidth]{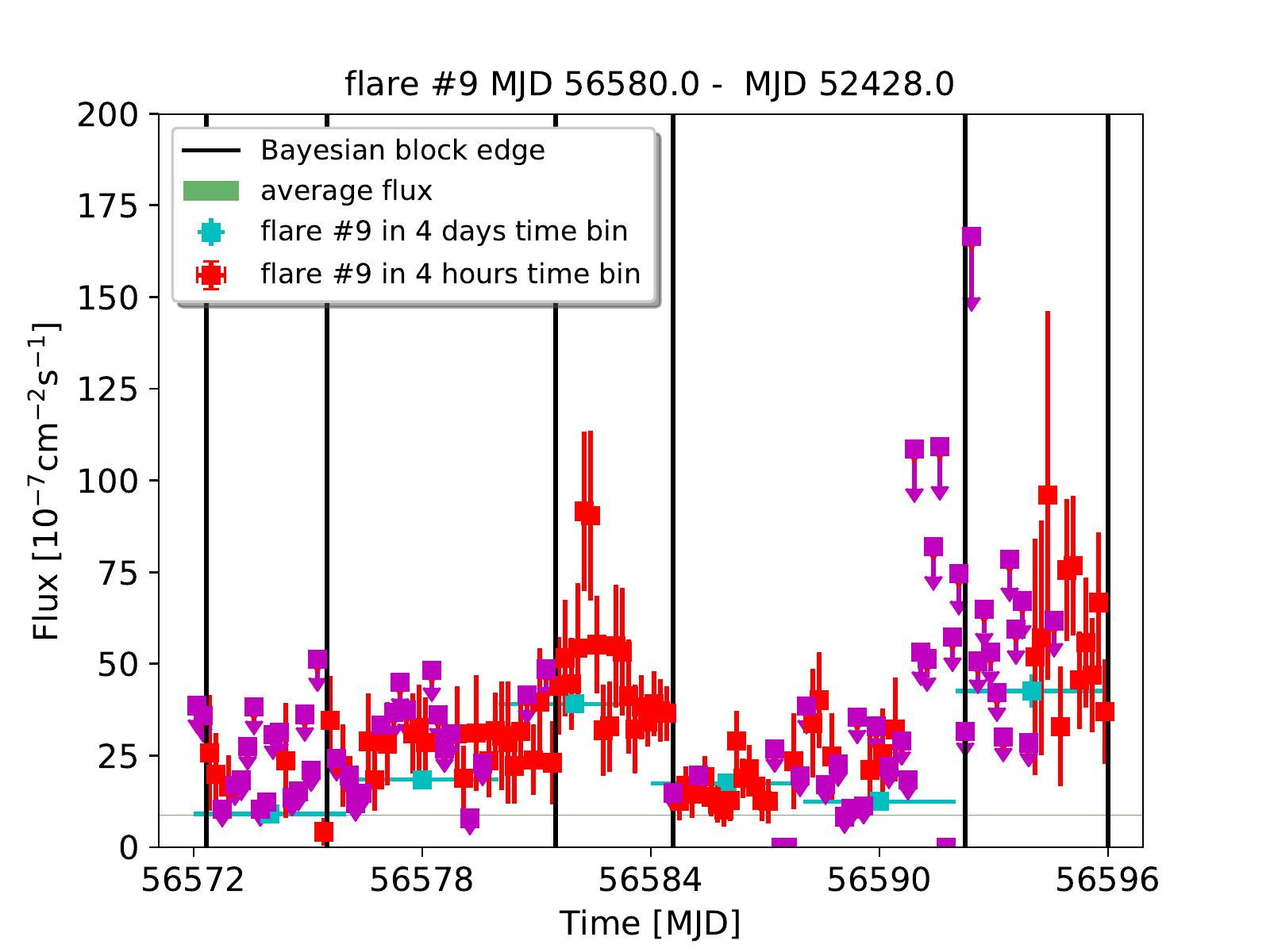}
\includegraphics[width=0.25\textwidth]{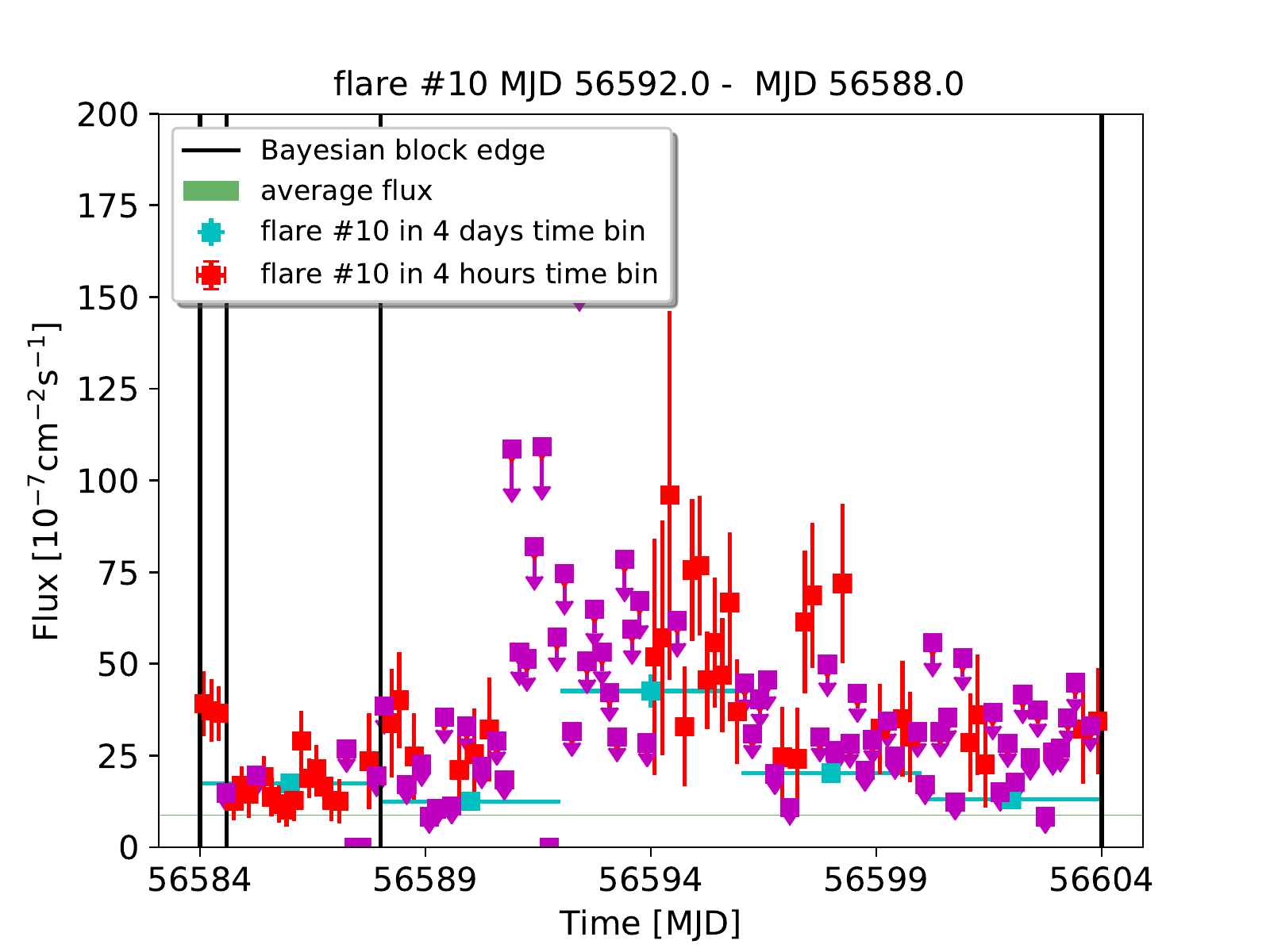}
\includegraphics[width=0.25\textwidth]{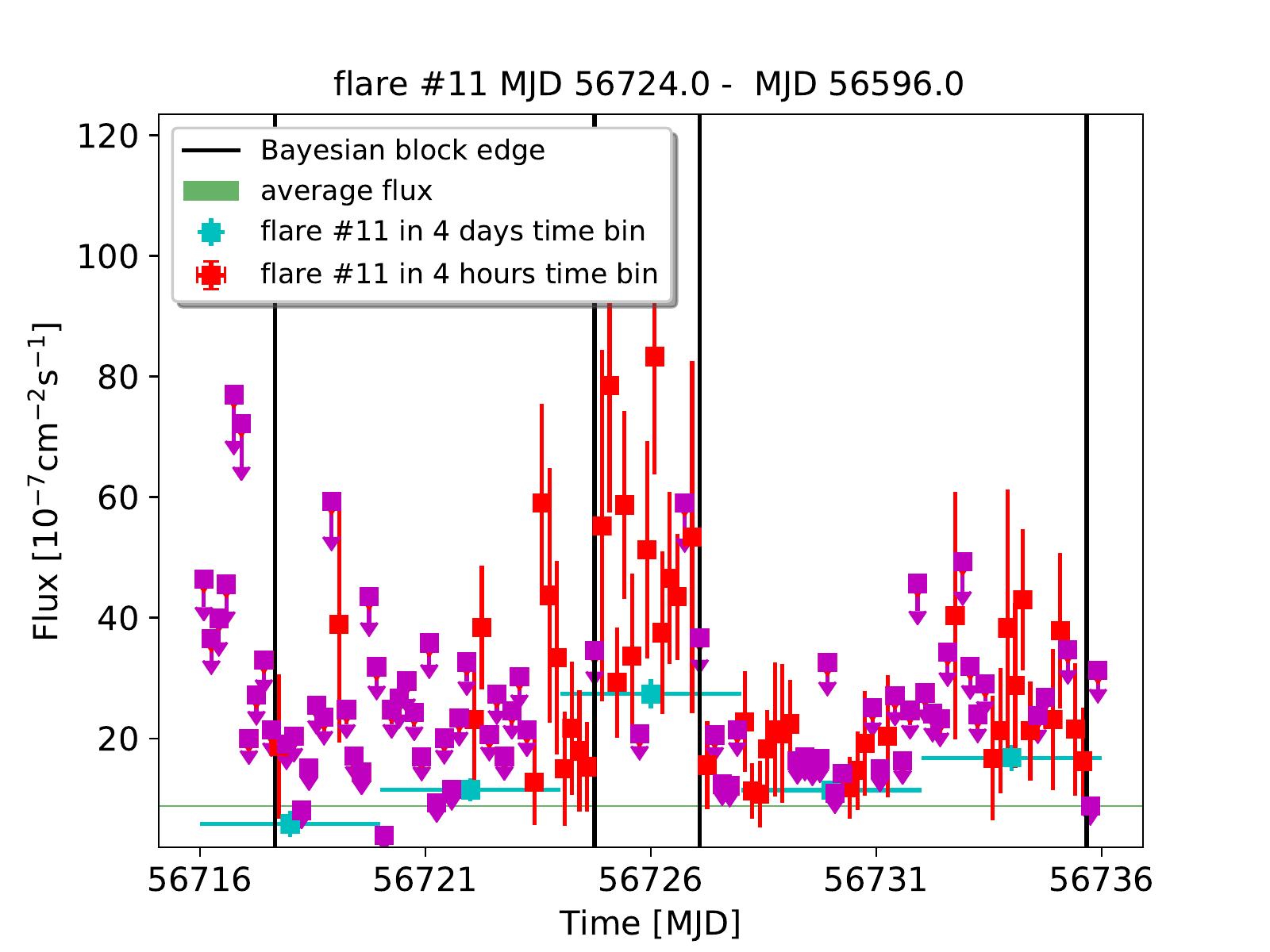}
\includegraphics[width=0.25\textwidth]{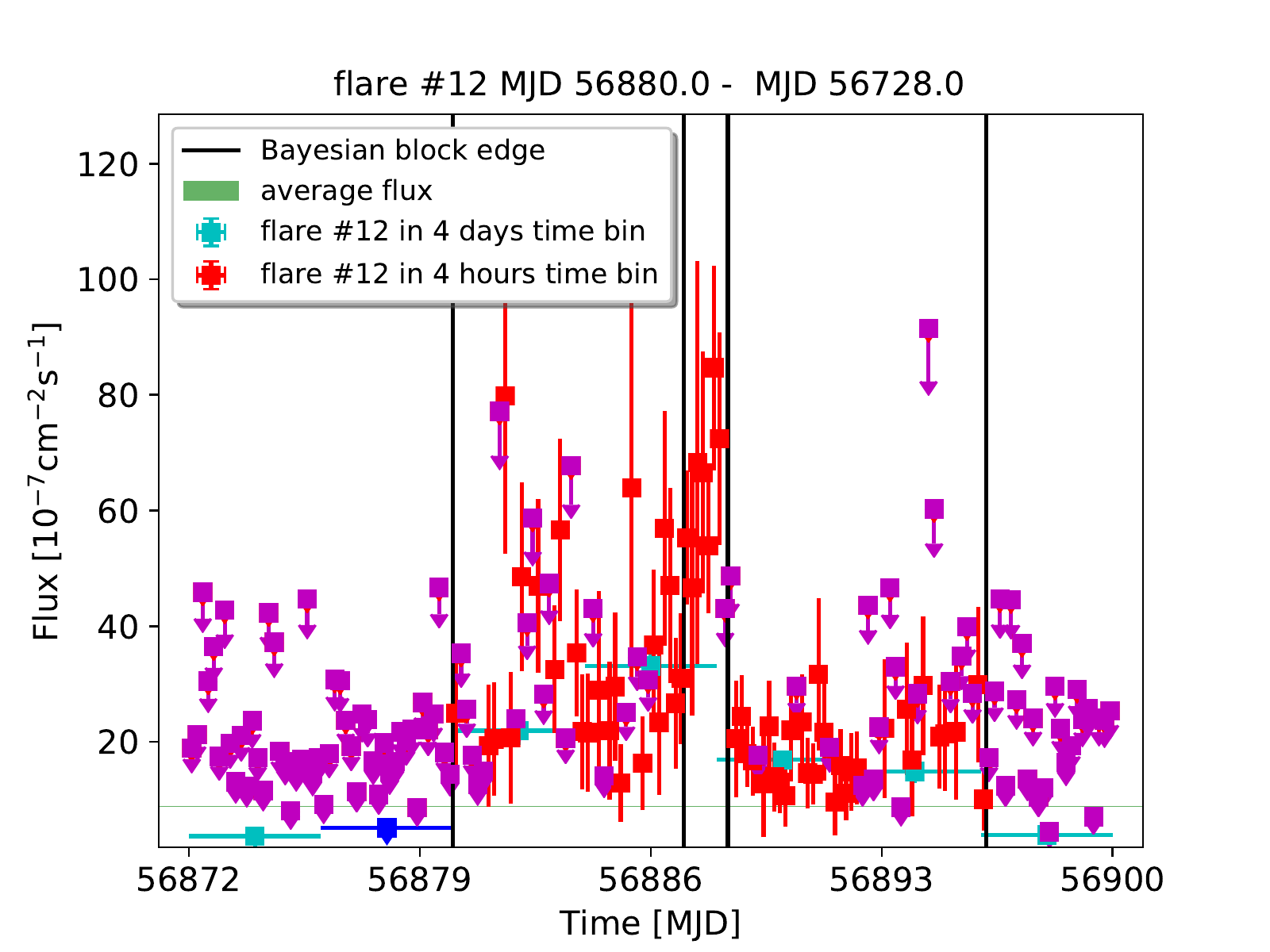}
\includegraphics[width=0.25\textwidth]{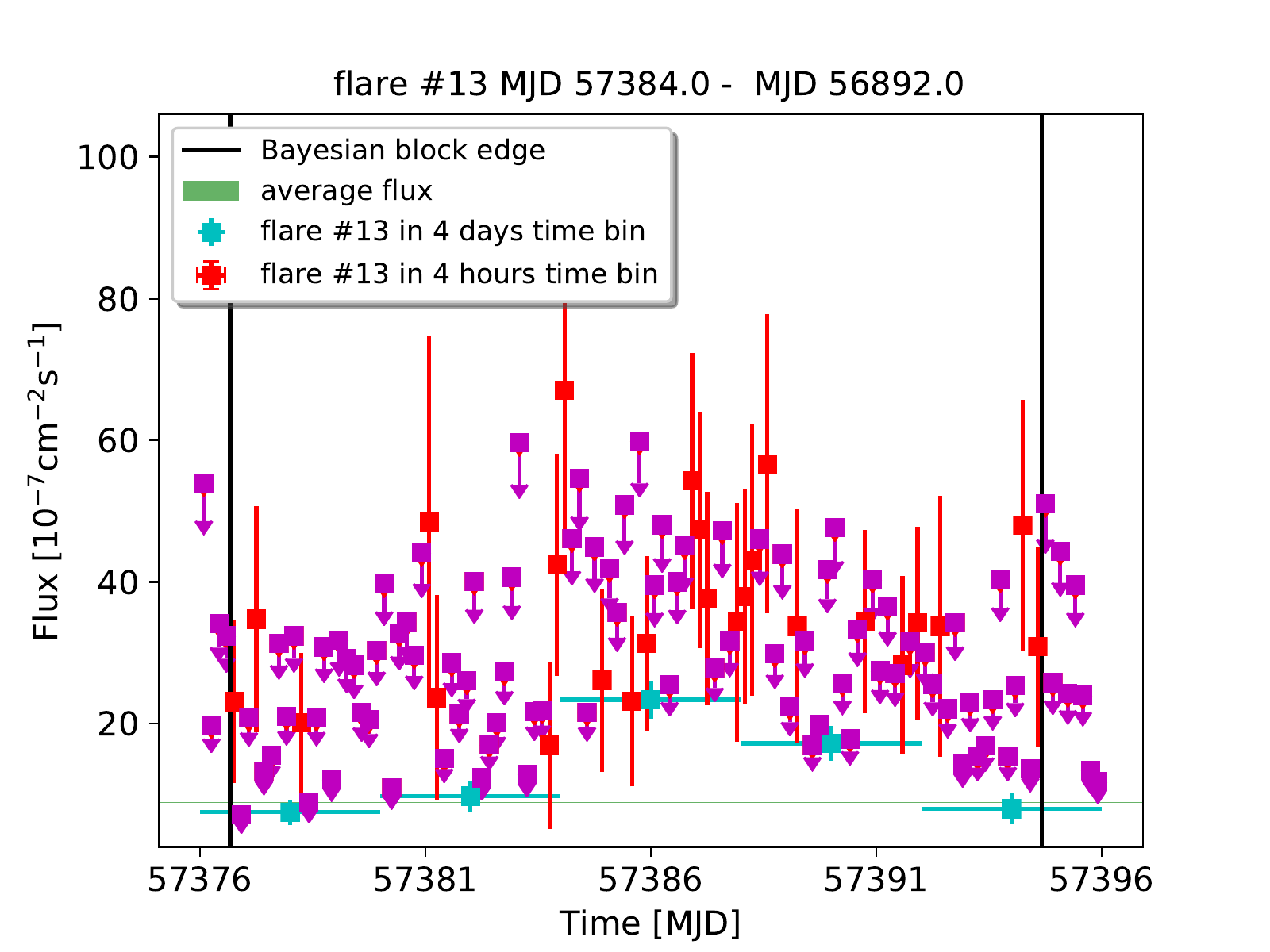}
\includegraphics[width=0.25\textwidth]{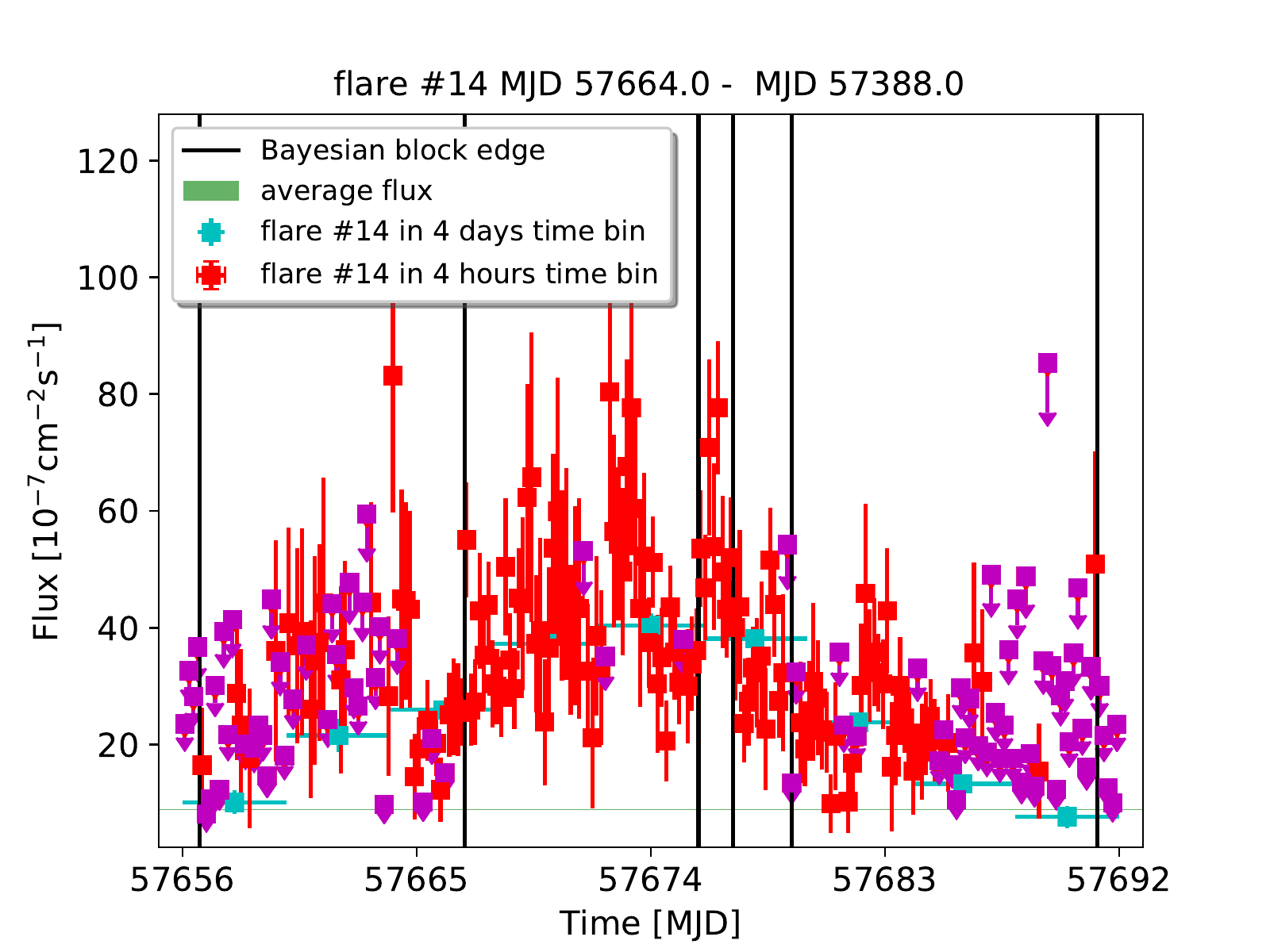}
\includegraphics[width=0.25\textwidth]{lc_flare_15_new3.pdf}
\includegraphics[width=0.25\textwidth]{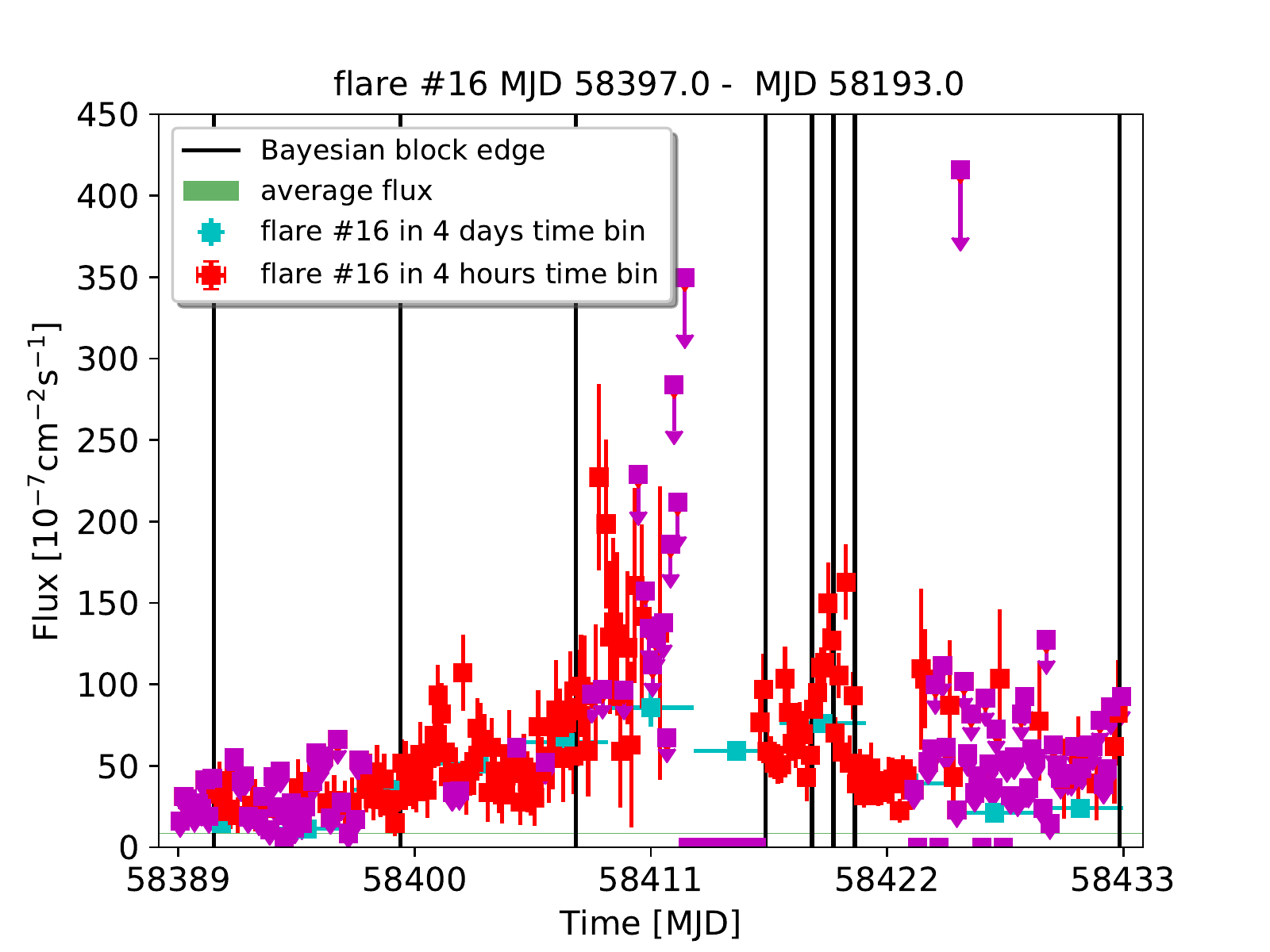}
\includegraphics[width=0.25\textwidth]{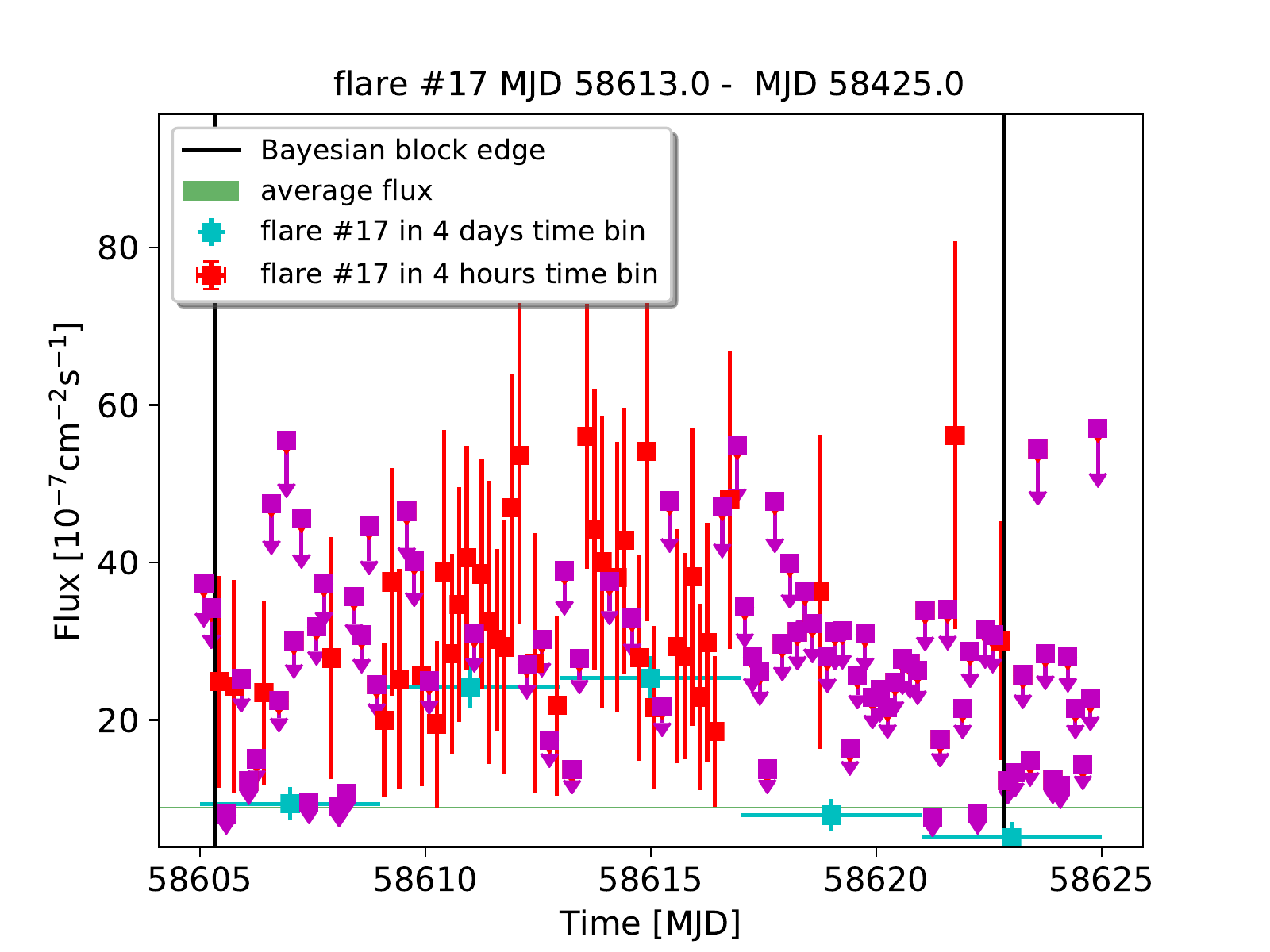}
\caption{Light curves of the low-energy component in 4-day binning, red for measured points and magenta for upper limits, and 4-hour binning, cyan for measured points and blue for upper limits,  for the 17 detected flares in 11 year observations of the Fermi-LAT. Edges defined by Bayesian block are shown in black vertical lines.  Average flux of the low-energy component in the whole observational time is in green band.
\label{fig:lc_bayes_all_flares}}

\end{figure*}

To define a flare, there are some sophisticated statistical methods in the literature, including the Bayesian block method \citep{2013ApJ...764..167S}, the sequential analysis and change point analysis \citep{Tartakovsky2020}, autoregressive modeling \citep{Box2015,2018FrP.....6...80F,Hyndman2018,Chatfield2019} and template fitting \citep{2016ApJ...829...23D}. But here we use a straight forward method to identify significant flares by selecting time bins with $5\sigma$ higher fluxes 
compared with the average one, $(8.45 \pm 0.08)\times 10^{-7}$
cm$^{-2}$~s$^{-1}$. If there are several times are adjacent, they are
defined as a single flare. We also try to use the Bayesian block method 
to detect flares \citep{2013ApJ...764..167S}. However, since the 
bin-by-bin variation is too large, the Bayesian block method does
not work efficiently for this light curve with 11 years data, and we only use it to detect flux variation of these identified flares in the following sections. For the 11 years of the Fermi-LAT data, we find 
17 such flares distributed in 35 time bins. All the flares reported before are detected using the above method. They are flare \#1 \citep{2011Sci...331..739A}, flare \#2 \citep{2011Sci...331..736T, 2011Sci...331..739A}, flare \#3 \citep{2012ApJ...749...26B, 2011ApJ...741L...5S}, and flare \#6 \citep{2013ApJ...775L..37M}.
Light curves and outburst times of the 17 flares are shown in Fig. ~\ref{fig:lc_bayes_all_flares}.

\section{The October 2018 flare}\label{sec:2018}

\begin{figure}
\centering
\includegraphics[width=0.48\textwidth]{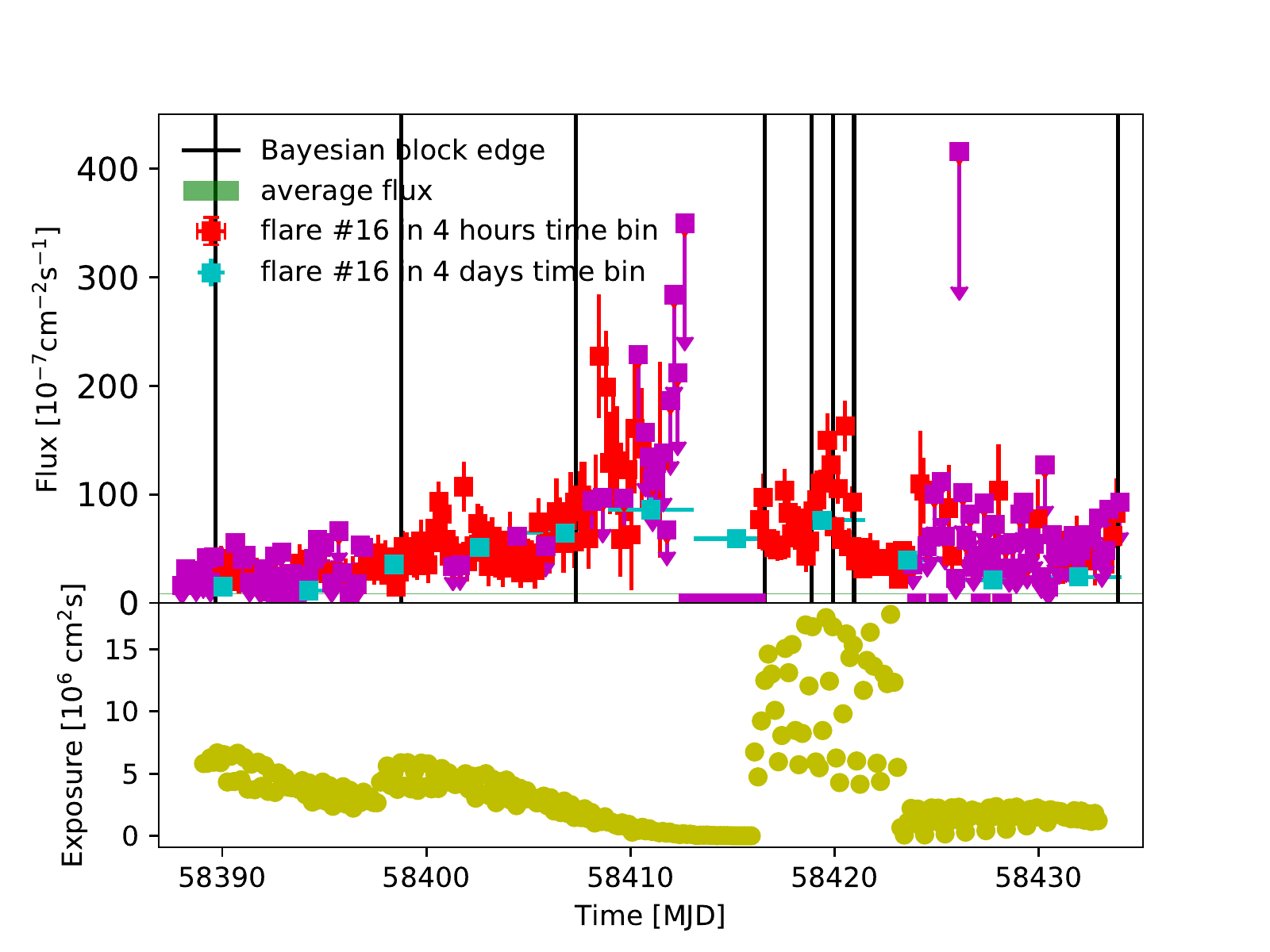}
\caption{Top: Light curves of the low-energy component in 4-day binning red for measured points and magenta for upper limits, and 
4-hour binning, cyan for measured points,  for the October 2018 flare. Edges defined by Bayesian block are shown in black vertical lines. Average flux of the low-energy component in the whole observational time is in green band. Bottom: The Exposure at the position of the Crab for every 4-hour observation around 100 MeV. 
\label{fig:flux_2018}}
\end{figure}

As can be seen in Fig. ~\ref{fig:flux_4days}, there is a very-long-duration flare 
occurred in October, 2018 (flare \#16), whose duration is about one month. 
This is by far the longest duration flare reported for the Crab Nebula. 
The light curve, in 4-day binning, of the $\gamma$-ray fluxes of this flare is shown in 
Fig.~\ref{fig:flux_2018}, where we add two more bins (8 days of data) before (after) the first (last) bin of the flare to easily see the rise of flare from the steady state. To get more details about this flare, we make the light curve with a bin width of 4 hours and use the Bayesian block method \citep{2013ApJ...764..167S}, with
a false positive rate of 0.07 \footnote{We test the false positive rate from 0.05 to 0.1, and determined time windows will not change.}, to determine the time windows, which contain information about variations of the flux. It seems that there are two sub-flares, peaking around MJD 58408 and MJD 58419. The rise of the first sub-flare is very slow, taking more than ten days to reach its peak. However, since there is not enough data between MJD 58411 and MJD 58416, when the exposure at the position of the Crab Nebula is very low or even zero as in the bottom panel of Fig. \ref{fig:flux_2018}, we can not get enough information about the decay of the first sub-flare and the rise of the second sub-flare. The decay time is approximately 1.5 days for the second sub-flare. A second rise or “shoulder”, around MJD 58410 and MJD 58420 with low significance, could be seen by eyes during the decaying phase of both sub-flares, similar to the flare happened in the April of 2011 \citep{2012ApJ...749...26B}, but the second rise of the second sub-flare has almost the same or even higher flux as its first rise.   

\begin{figure}
\centering
\includegraphics[width=0.48\textwidth]{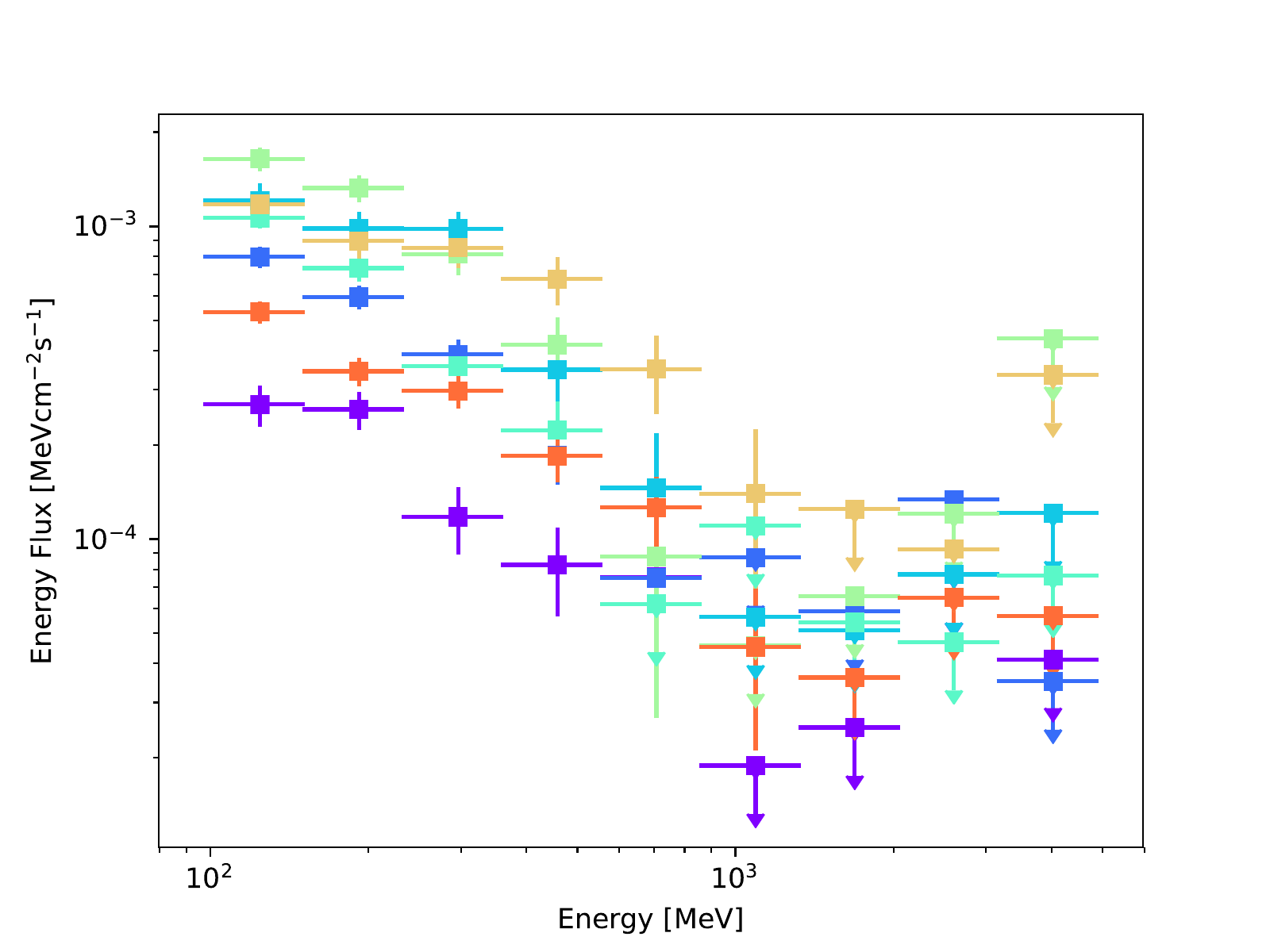}
\caption{The SEDs of the 7 time bins, defined by the Bayesian block method, for the low-energy component of the October 2018 flare.
\label{fig:sed_flares_2018}}
\end{figure}

\begin{figure}
\centering
\includegraphics[width=0.48\textwidth]{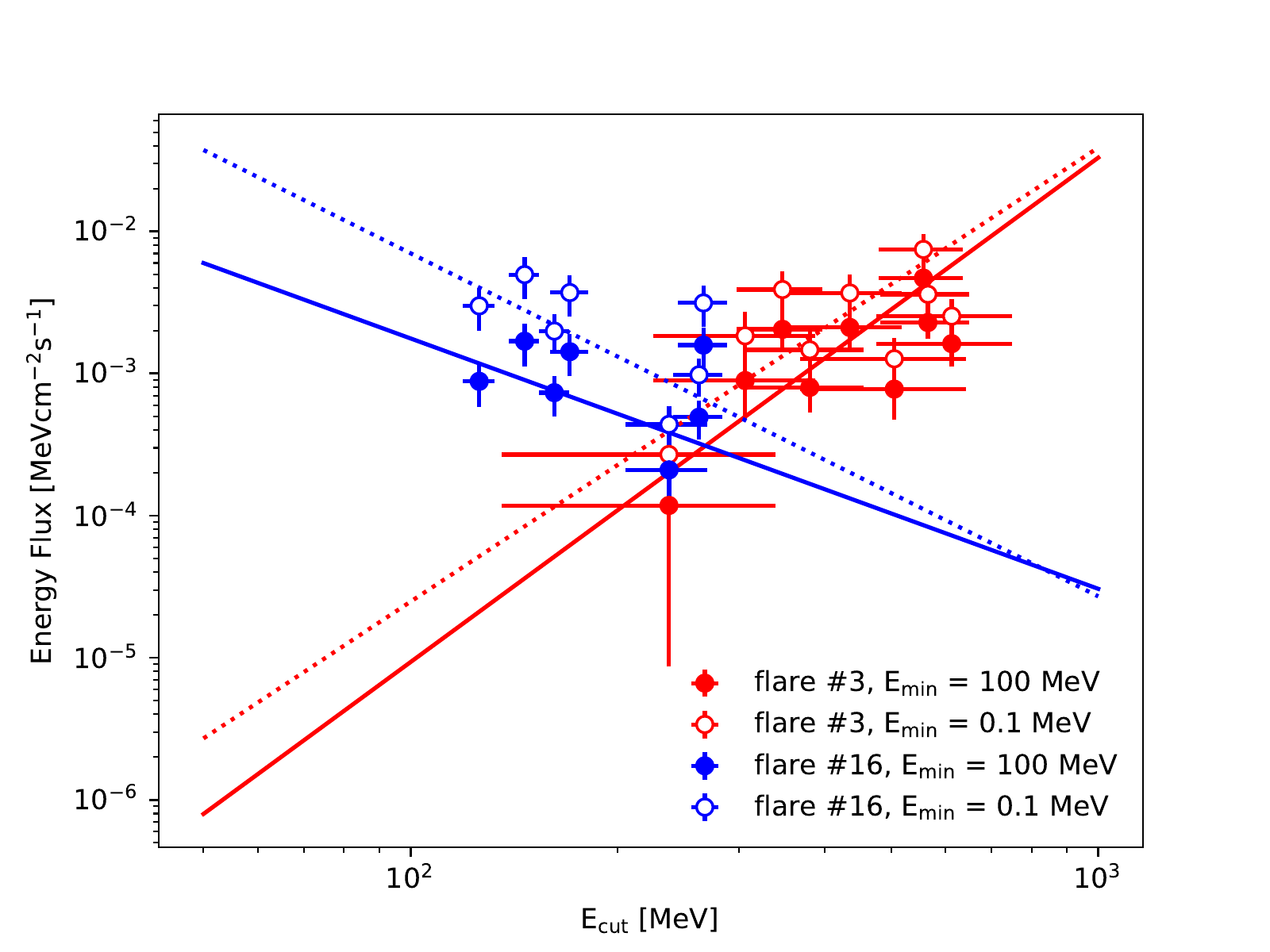}
\caption{The energy flux versus the cutoff energy of the flare  component for flare \#3 and flare \#16. Two minimum energies are used to get those energy fluxes. And a PL function is used to fit the relationship.
\label{fig:EcutEF_2018}}
\end{figure}

For the 7 time bins defined by the Bayesian block method, we derive their spectral energy distributions (SED) of the low-energy component, through dividing the data into 9 logarithmically evenly distributed energy bins from 100 MeV to 5 GeV to cover the energy range of the low-energy component, for each time bin. Only the normalizations of the low-energy component are set free during the SED fits, while other parameters are fixed as best fit values evaluated in the previous section for the joint likelihood analysis of all time intervals. If the significance is lower than $2\sigma$ in particular energy bins, the 95\% upper limits are derived. The results are given in Fig.~\ref{fig:sed_flares_2018}. One SED, corresponding to the time window from MJD 58419.5 to MJD  58420.5, has data point reaching 1 GeV. 

Following \citet{2012ApJ...749...26B} we assume that the total emission 
of the low-energy component could be further composed of two components,
a steady background and a flare component, which is assumed to only contribute to the low energy part considering no firm detection at very-high energy yet \citep{2012ATel.4258....1B, 2014A&A...562L...4H,2014ApJ...781L..11A,2015JHEAp...5...30A}. A PL model,  $dN/dE \propto E^{-\Gamma_{b}}$, is adopted to describe 
the background emission, and a power-law with an exponential cutoff (PLEC),  $dN/dE \propto E^{-\Gamma_{f}}exp(-E/E_{cut})$, 
is used to describe the flare emission. The steady background component and the index $\Gamma_{f}$ of the flare component are assumed to be the same in all time windows, but the normalization and the cutoff energy $E_{\rm cut}$ are free parameters for each time window.  Using the module {\textit{Composite2}} 
in the Fermitools, we fit all the 7 time bins simultaneously. The integrated flux of the steady component is $\Phi_{100}=(9.53 \pm 2.06) 
\times 10^{-7}$~cm$^{-2}$~s$^{-1}$, and the spectral index is
$\Gamma_b=4.06 \pm 0.07$. The spectral index of the flare component, $\Gamma_f$, is
derived to be $1.33 \pm 0.25$. To check the validity of our method, we also use the same method to re-analyze the April 2011 flare (flare \#3), and we get $\Phi_{100}=(9.90 \pm 3.27) 
\times 10^{-7}$~cm$^{-2}$~s$^{-1}$, $\Gamma_b=3.53 \pm 0.28$ and $\Gamma_f$= $1.37 \pm 0.12$. All these results, especially the spectral index of the flare component of these two flares, are consistent with those derived for the April 2011 flare in \citet{2012ApJ...749...26B}, where they get $\Phi_{100}=(5.4 \pm 5.2) 
\times 10^{-7}$~cm$^{-2}$~s$^{-1}$, $\Gamma_b=3.9 \pm 1.3$ and $\Gamma_f$= $1.27 \pm 0.12$.

We plot in Fig.~\ref{fig:EcutEF_2018} the fitting results of the cutoff energies, $E_{\rm cut}$, and the energy flux, $F$, of the flare component for both flares, flare \#3 and flare \#16. Following \citet{2012ApJ...749...26B}, we use a PL function $F=\alpha E_{\rm cut}^{\beta}$ 
to fit the relationship between both quantities. We first set $E_{\rm min}$ to 100 MeV to get the energy flux. For flare \#3, we have $\beta = 3.55 \pm 1.37$, which is consistent with previous result \citep{2012ApJ...749...26B}. But for flare \#16, we get $\beta = -1.76 \pm 0.88$, which shows a different relationship between $E_{\rm cut}$ and $F$. For flare \#16, some of the cutoff energy are close to 100 MeV, and choosing 100 MeV as $E_{\rm min}$ to get the energy flux will lead to more underestimation of the energy associated with these low cutoff energy time window. Here we also test to switch $E_{\rm min}$ to 0.1 MeV, and as in Fig. ~\ref{fig:EcutEF_2018} this would make $\beta$ smaller for both flares and lead to larger deviation between $\beta$ derived for these two flares, with  $\beta = 3.20 \pm 1.31$ and  $\beta = -2.41 \pm 0.89$, respectively. 

Thus for this October 2018 flare (flare \#16), it could be decomposed into a steady background component, which has consistent parameters as the steady background component of flare \#3, and a flare component, which also has a consistent spectral index as the flare component of flare \#3. This indicates that these two flares may share similar mechanism behind them. However there may be a discrepancy between indexes $\beta$, which shows the relationship between the energy flux and the cutoff energy of the flare component, derived from these two flares, and this may lead to more complicated modelling for the emission mechanism.

\section{Properties of the flares}\label{sec:combined}

As shown in Section \ref{sec:2018}, flares occurred in April, 2011 and October, 2018 may share the same steady background and the same spectral index for the flare component, thus we may expect that all 17 flares will share these parameters, only varying the normalization and cutoff energy to account for their difference. Then the same as for flare \#16, we make light curves with a bin width of 4 hours for each flare identified in Section \ref{sec:flares}, and use the Bayesian block method to determine time windows, as shown in Fig. ~\ref{fig:lc_bayes_all_flares}.  Using the module {\textit{Composite2}} 
in the Fermitools, we fit the normalization and the cutoff energy of flare component for each time window, spectral index for the flare component and parameters for the steady background component simultaneously. This composite likelihood analysis gives an integrated flux of the steady
background component $\Phi_{100}=(8.22 \pm 1.14) \times 10^{-7}$
cm$^{-2}$~s$^{-1}$ above 100 MeV, a spectral index $\Gamma_b=3.80\pm0.15$, and a spectral index of the flare component $\Gamma_f=1.39 \pm 0.08$. As expected, these parameters, derived by combining all identified flares, are consistent with those derived from analysis for individual flare, and this may indicate all flares would share the same emission mechanism.    

\begin{figure}
\centering
\includegraphics[width=0.48\textwidth]{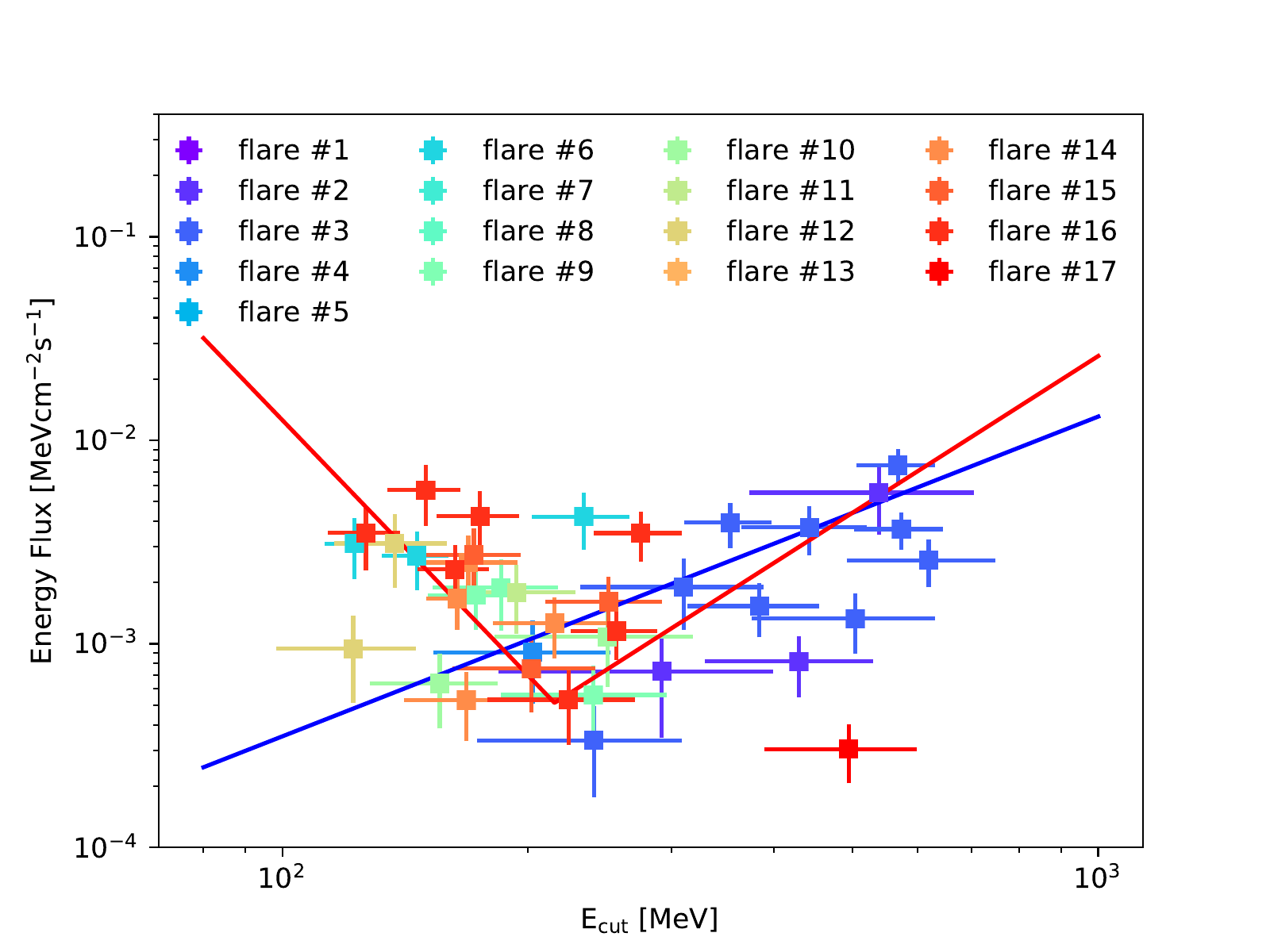}
\caption{The energy flux versus the cutoff energy of the flare  component for all identified flares together. Different colors are used to indicate points associated with different flares. 0.1 MeV is used as the minimum energy to get those energy fluxes. A PL function and a broken power-law are fitted, and are shown in blue and red line. 
\label{fig:EcutEF}}
\end{figure}

We plot, in Fig.~\ref{fig:EcutEF}, the fitting results of the cutoff energies and the energy flux, with $E_{\rm min}= 0.1$ MeV, of the flare component for all time windows associated with identified flares, excluding those with TS $<$ 25 and those with $E_{\rm cut}$ $<$ 100 MeV, which may be not so reliable considering our data are all above 100 MeV. Using a PL function $F=\alpha E_{\rm cut}^{\beta}$ to fit the results gives $\beta = 1.57 \pm 0.33$, with a reduced $\chi^2$ 
of about 3.03 for a number of degrees of freedom (dof) 37. Motivated by possibly different $\beta$ for flare \#3 and flare \#16, we use a broken power-law, $F=\alpha (E_{\rm cut}/E_{\rm break})^{\beta_{1}}$ for $E_{\rm cut}<E_{\rm break}$ and  $F=\alpha (E_{\rm cut}/E_{\rm break})^{\beta_{2}}$ for $E_{\rm cut}>E_{\rm break}$, to fit the relationship between the cutoff energy and energy flux. This would give a better fit, compared with the PL function, with a reduced $\chi^2$ of about 1.88 for a number of dof 35. And this would give a break energy, $215.50 \pm 19.00$ MeV, the index $\beta_{1}=-4.16 \pm 1.32$, and the index $\beta_{2}=2.56 \pm 0.61$. And there is a marginal consistency, between $\beta_{1}$ and $\beta$ derived for flare \#16, and between $\beta_{2}$ and $\beta$ derived for flare \#3. These may suggest there are two groups of flares, as hinted in Section \ref{sec:2018}, and some mechanisms, other than Doppler boosting, may be needed.

\begin{figure}
\centering
\includegraphics[width=0.48\textwidth]{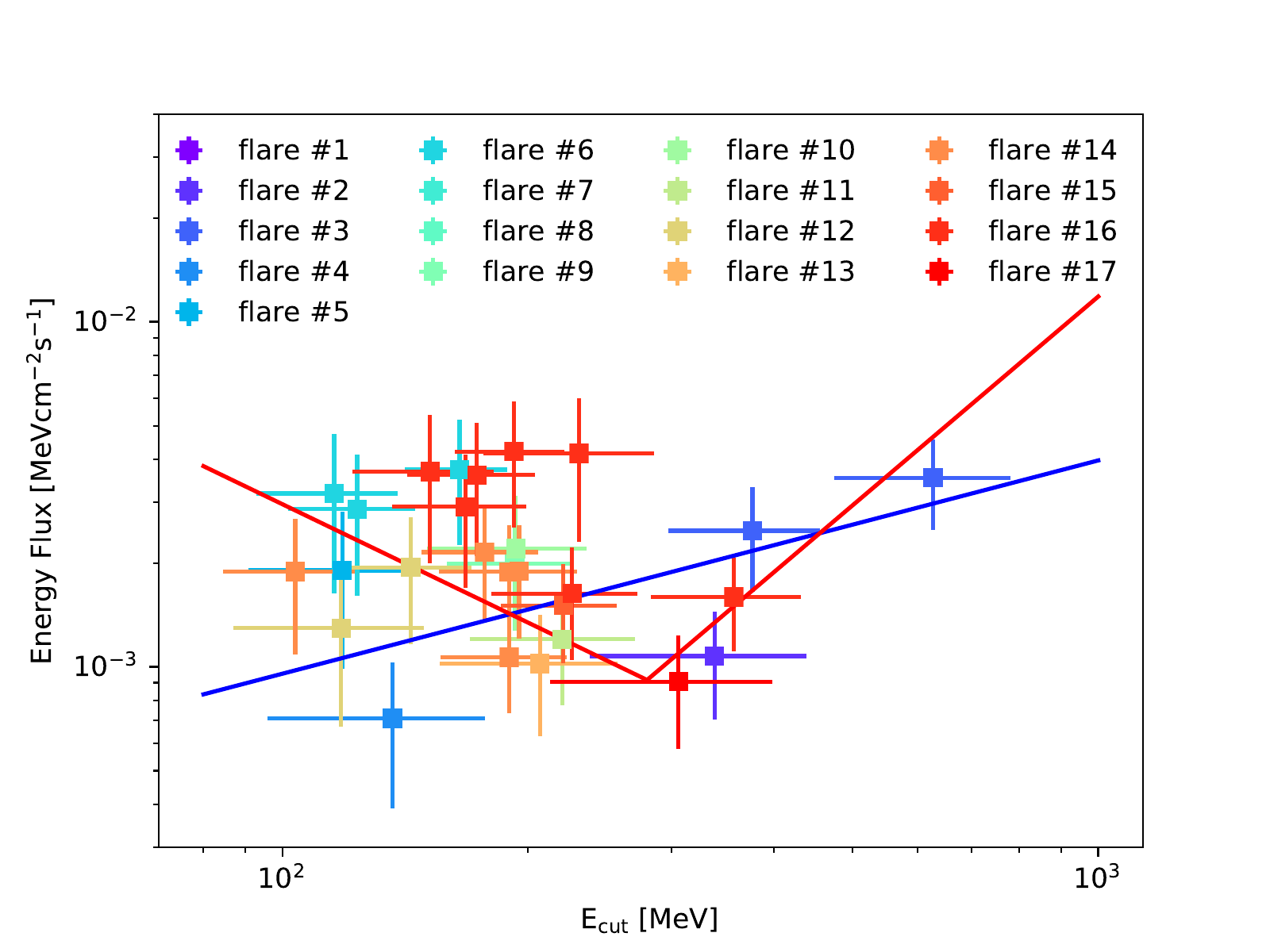}
\caption{Same as Fig.~\ref{fig:EcutEF}, but the time bin is fixed to 4-days for all flares. 
\label{fig:EcutEF_4days}}
\end{figure}

To further investigate the possible effect by the width of the time 
bin, we repeat all the analyses with time bin widths fixed to 4 days. And we get the integrated flux of the steady background component $\Phi_{100}=(6.55 \pm 3.34) \times 10^{-7}$
cm$^{-2}$~s$^{-1}$ above 100 MeV, a spectral index  $\Gamma_b=3.55\pm0.25$, and a spectral index of the flare component $\Gamma_f=1.52 \pm 0.22$, which are all consistent with those derived previously. A PL function and a broken power-law are used to fit the relationship between the cutoff energy and the energy flux of each time bin, as shown in Fig. \ref{fig:EcutEF_4days}, and still a broken power-law, with one index negative and another index positive, is weakly preferred.

\begin{table*}
\centering
\scriptsize
\caption{Parameters about the flux above 100 MeV and the index of the steady background, the index of the flare component and parameters for the relationship between the cutoff energy and the energy flux above 0.1 MeV of the flare component. Parameters from individual flare, flare \#16 and flare \#3, and from all flares together are shown. For comparison, we also list parameters for flare \#3 from \citep{2012ApJ...749...26B}, where they consider energy flux above 100 MeV to get $\beta$, and we list parameters derived for all flares, with fixed 4-day binning .} 
\begin{tabular}{|c|c|c|c|c|c|}
\hline
 & flare \#16 (Bayesian block)& flare \#3 (Bayesian block)& flare \#3 \citep{2012ApJ...749...26B} & all flares (Bayesian block) & all flares (4-day binning) \\ \hline
 $\Phi_{100}$ $(10^{-7}$~cm$^{-2}$~s$^{-1})$ &  9.53 $\pm$ 2.06          &  9.90 $\pm$ 3.27         &   5.4 $\pm$ 5.2        & 8.22 $\pm$ 1.14 & 6.55 $\pm$ 3.34 \\ \hline
$\Gamma_{b}$ &   4.06 $\pm$ 0.07         &  3.53 $\pm$ 0.28         &    3.9 $\pm$ 1.3        & 3.80 $\pm$ 0.15   & 3.55 $\pm$ 0.25 \\ \hline
$\Gamma_{f}$ &  1.33 $\pm$ 0.25          &   1.37 $\pm$ 0.12          &   1.27 $\pm$ 0.12        & 1.39 $\pm$ 0.08 & 1.52 $\pm$ 0.22 \\ \hline
$\beta$&  -2.41 $\pm$ 0.89          &   3.20 $\pm$ 1.31          &   3.42 $\pm$ 0.86        & 1.57 $\pm$ 0.33 & 0.62 $\pm$ 0.25 \\ \hline
$\beta_{1}$, $\beta_{2}$ &            &            &           & -4.16 $\pm$ 1.32, 2.56 $\pm$ 0.61 & -1.14 $\pm$ 0.48, 2.01 $\pm$ 1.27 \\ \hline
\end{tabular}
\label{Tab:decomposition}
\end{table*}

\section{Discussion}\label{sec:discussion}

\begin{figure}
\includegraphics[width=0.48\textwidth]{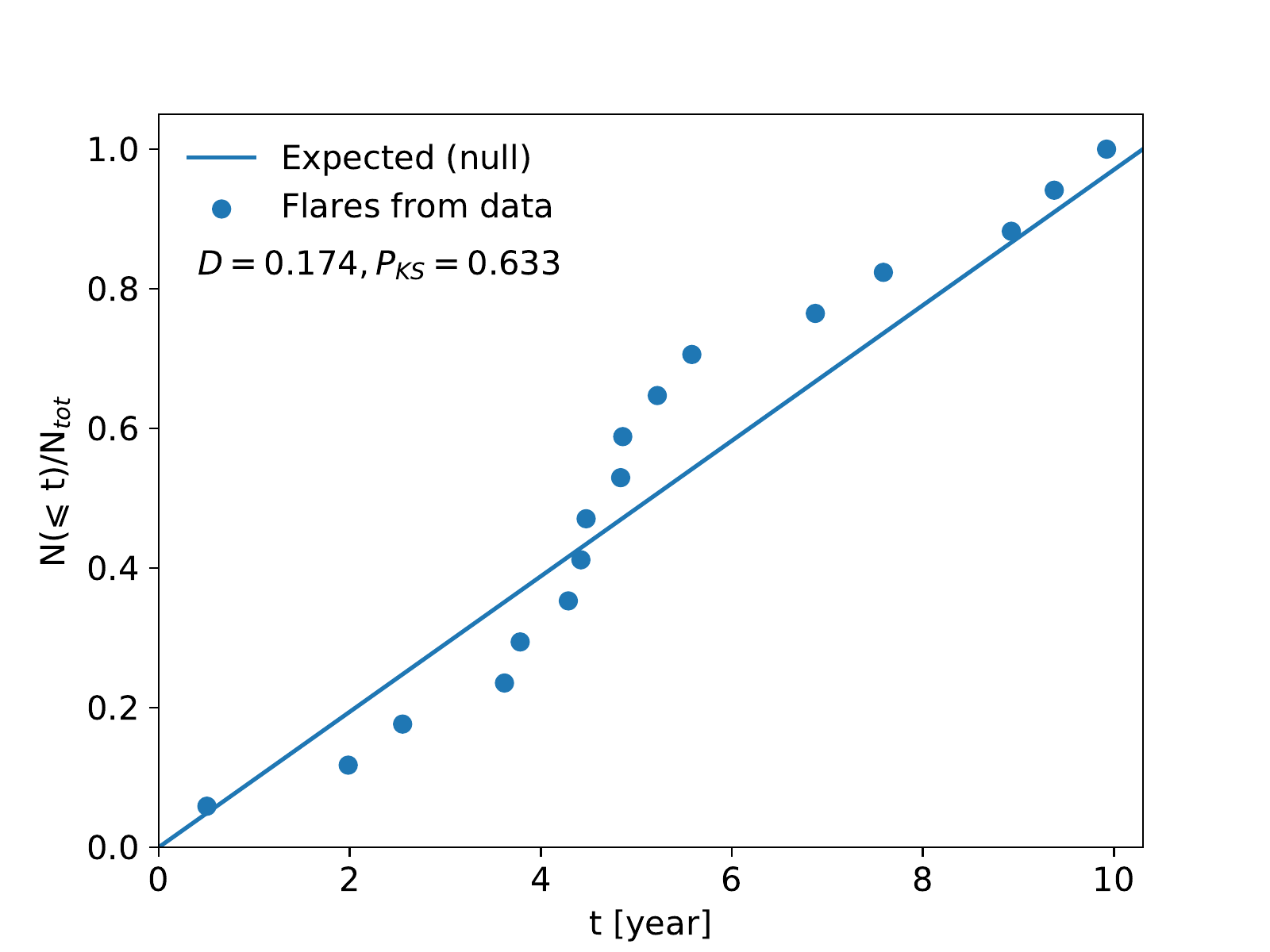}
\includegraphics[width=0.48\textwidth]{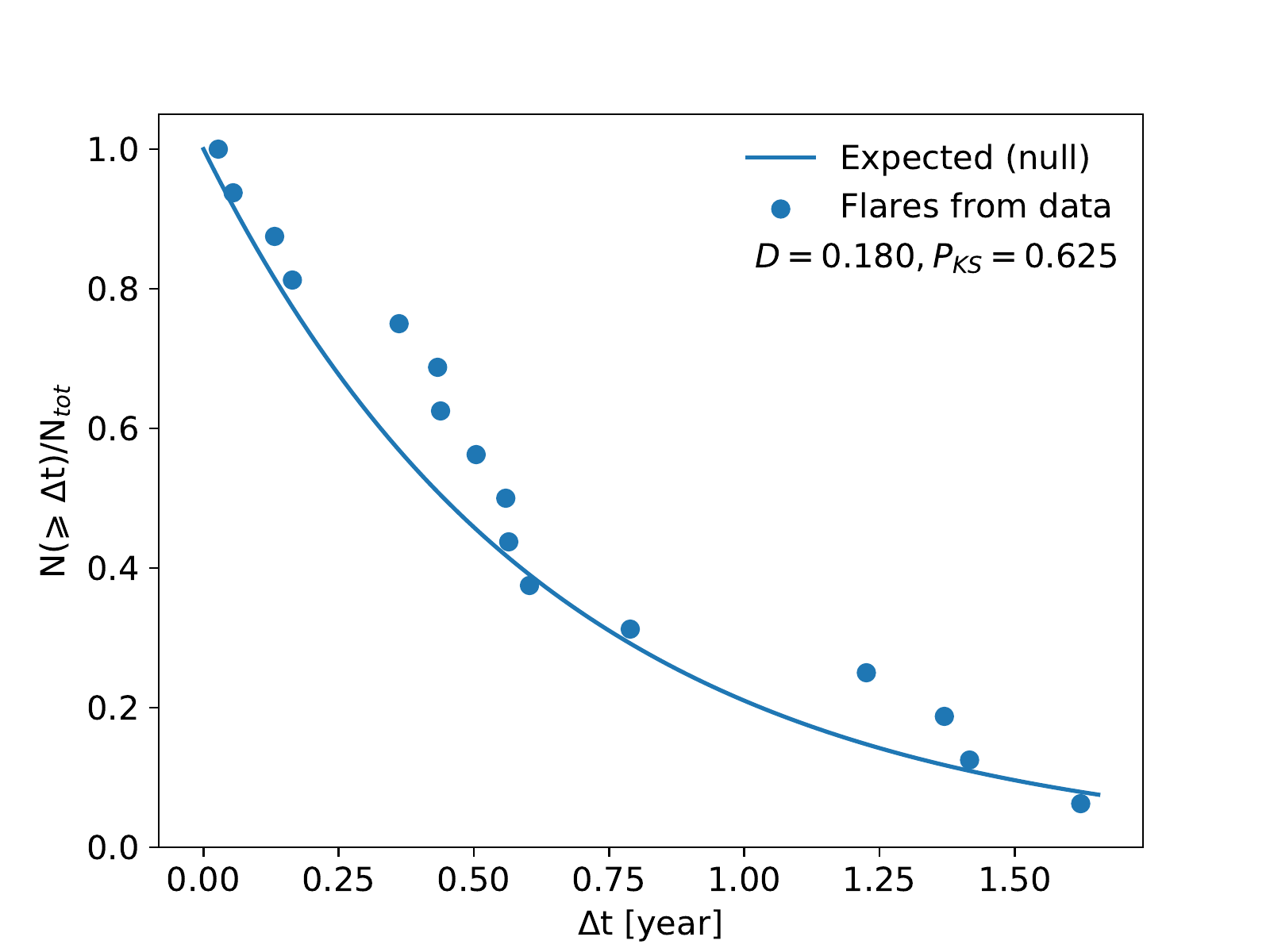}
\caption{Normalized cumulative number of flares as a function of time (top) and the normalized cumulative distribution of the waiting time (bottom).    
\label{fig:rate}}
\end{figure}

In the 11 years of the observational data, we identify 17 significant 
flares, which corresponds to a flare rate of $\sim1.5$ yr$^{-1}$. 
It would be interesting to explore the potential long-term change of 
the flare rate, and to search for possible clustering of the flares
\citep{2016MNRAS.456.1438Y}. We calculate the cumulative number of flares 
as a function of time. The nearly linear increase of the number with time, 
as shown in the top panel of Fig.~\ref{fig:rate}, suggests that the flare 
rate is approximately constant during the observations. 
The Kolmogorov-Smirnov (KS) test of the data versus the constant expectation
gives a probability of $P_{\rm KS}=0.633$. We also calculate the cumulative 
distribution of the waiting time between successive flares, as given in the
bottom panel of Fig.~\ref{fig:rate}. Compared with the null hypothesis in 
which flares occur randomly with an exponential distribution of the waiting 
time, $dN/d(\Delta t) \propto \exp(-\Delta t/\overline{\Delta t})$ with
$\overline{\Delta t}$ being the mean waiting time from the 17 detected 
flares, the occur rate of flares is consistent with a stationary Poisson 
process with a constant rate. The current data do not show any significant 
clustering of the flare rate.

Our work shows that not only individual flare, but all flares together, could be decomposed into a steady background component and a varying flare component. And as summarized in Table \ref{Tab:decomposition}, the flux and the index $\Gamma_{b}$ of the background component, and the index $\Gamma_{f}$ of the flare component are consistent for all cases. This may be a strong indication that all flares would share the same emission mechanism. But for the relationship between the cutoff energy and the energy flux, there is a break between 200 to 300 MeV.  As hinted by colors for different flares in Fig. \ref{fig:EcutEF}, there may be two groups of flares, one, such as flare \#3, with large cutoff energies and a positive index for a PL function, and the other, such as flare \#16, with low cutoff energies and a negative index for a PL function \footnote{We make the same analysis for a sub-group of flares, excluding flare \#2 and flare \#3 , which contribute to most of points with large $E_{cut}$ in Fig. \ref{fig:EcutEF}. Again we get consistent parameters for the steady background and the index of the flare component. A power-law relation with a negative index would be achieved, consistent with that for the flare in October, 2018.}. The positive index, which is about 3, is consistent with that  expected for the Doppler boosting scenario of the flare emission. But the physical interpretation for the negative index would need investigation in future works. We also note that the goodness for the fitting of the relationship is not good, even for the broken power-law case, with a reduced $\chi^2$ of about 1.88 for a number of dof 35. This may be due to the intrinsic dispersion of different flares.

\section{Conclusion}\label{sec:conclusion}

Energetic flares have been revealed in the $\gamma$-ray band from the Crab 
Nebula by AGILE and Fermi-LAT \citep{2011Sci...331..736T,2011Sci...331..739A,
2012ApJ...749...26B,2011ApJ...741L...5S,2013ApJ...775L..37M}. 
Using the eleven years of the Fermi-LAT data, we carry out a systematic
search for flares from the Crab Nebula in this work. We confirm that the 
flux of high-energy component was stable with deviations from the average 
flux $<3\sigma$. The low-energy component is found to be highly variable.
We identify 17 significant flares from 2008 to 2019, which correspond to 
a flare rate of $\sim1.5$ yr$^{-1}$. The data is consistent with a
random occur rate of flares, without significant clustering of the flares. 

We have done a case study for the longest duration flare occurred in October, 2018. We have a detection of synchrotron photons up to energies of about 1 GeV. And we also find it could be decomposed in a similar way, with a steady soft PL component and a variable PLEC flaring component, as the strong flare occurred in April, 2011, with consistent parameters. But the relationship between the cutoff energy and the energy flux is different from that for the flare occurred in April, 2011, with a negative index. 

We make a similar decomposition for all identified flares as we did for individual flare, and we get a consistent description for the steady background component and the index of the flare component. In this case, the steady component has a PL index of $3.80\pm0.15$, which may correspond
to the synchrotron tail of the accelerated electrons of the overall
nebula. The PL index of the flare component is much harder, 
$\Gamma=1.39\pm0.08$, with an exponential cutoff which varies flare
by flare (or even bin by bin for the same flare).  This consistent picture may suggest that all flares would share similar emission mechanism.  And the cutoff energies
often, about 82\% of points in the Fig.~\ref{fig:EcutEF}, exceed the $\sim160$ MeV synchrotron limit of diffusive shock
acceleration models, indicating the existence of Doppler boosting
or special acceleration mechanisms \citep{2011MNRAS.414.2017K,
2011ApJ...730L..15Y,2012MNRAS.424.2249K,2011ApJ...737L..40U}. But we also find that, compared with a PL, a broken power-law could better fit the relationship between the cutoff energy and the energy flux of these flares. And for the best fitted broken power-law, one index is consistent with that derived for the flare occurred in October, 2018 and another index is consistent with that derived for the flare occurred in  April, 2011. While the positive index around 3 is consistent with the Doppler boosting scenario of the flare emission, but the negative index needs further investigation.

Finally we emphasize that the statistical properties of the detected
flares should be useful in understanding the physical mechanism of
the flare production. A dedicated statistical study with template fitting method for flare detection, with a focus on 
the flare energy distribution, duration distribution, and the 
energy-duration correlation, will be published elsewhere.

\section*{Acknowledgments}
We acknowledge the use of the Fermi-LAT data provided by the Fermi Science 
Support Center. 
This work is supported by the National Key Research and Development Program
(No. 2016YFA0400200), the National Natural Science Foundation of China 
(Nos. 11525313, 11722328, 11851305), and the 100 Talents program of Chinese 
Academy of Sciences. QY is also supported by the Program for Innovative 
Talents and Entrepreneur in Jiangsu.

\bibliographystyle{aasjournal}
\bibliography{refs}

\begin{appendix}
\section{Photon Selection with different pulsar phases}\label{sec:phasecut}
\begin{figure*}[!htb]
\centering
\includegraphics[width=0.8\textwidth]{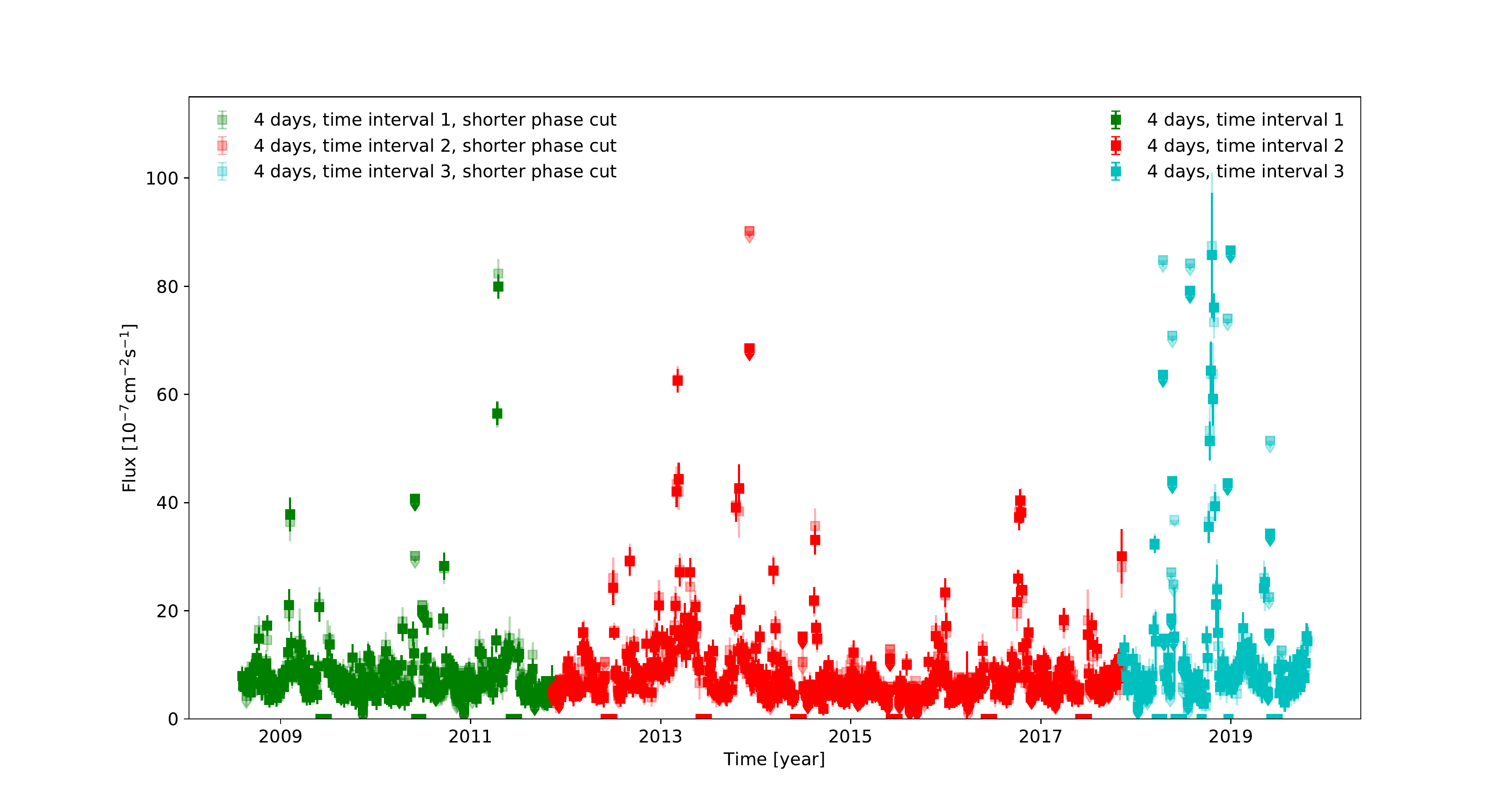}
\caption{Same as Fig.~\ref{fig:flux_4days} but overploted with the results
for more strict phase cuts.
\label{fig:flux_4days_shortcut}}
\end{figure*}

To check the effect of the photon selection with different pulsar
rotational phases, we follow the procedure in Section \ref{sec:flares} 
but choose photons with narrower phases, from 0.29 to 0.59, from 0.61 to 0.91, 
and from 0.45 to 0.75 for the three time intervals, respectively. 
The light curve of the low-energy component is shown in Fig. 
\ref{fig:flux_4days_shortcut}. For almost all time bins, the fluxes 
derived from data with shorter phase cuts are consistent with that 
derived from data with original phase cuts. The results of our analysis
should be affected little by the phase cuts.

\end{appendix}


\end{document}